\documentclass[twocolumn,superscriptaddress,altaffilletter]{revtex4-1}
\usepackage{graphicx,amsmath,amssymb}

\begin{document}

\title{Tunable Ultrafast Thermal Relaxation in Graphene \\Measured by Continuous-Wave Photomixing}

\author{M. Mehdi Jadidi}
\email{mmjadidi@umd.edu}
\affiliation{Institute for Research in Electronics \& Applied Physics, University of Maryland, College Park, MD 20742, USA}

\author{Ryan J. Suess}
%\email{}
\affiliation{Institute for Research in Electronics \& Applied Physics, University of Maryland, College Park, MD 20742, USA}

\author{Cheng Tan}
\affiliation{Department of Mechanical Engineering, Columbia University, New York, New York 10027, USA}

\author{Xinghan Cai}
%\email{}
\affiliation{Department of Physics, University of Washington, Seattle, Washington 98195, USA}

\author{Kenji Watanabe}
\affiliation{National Institute for Materials Science, 1-1 Namiki, Tsukuba 305-0044, Japan}

\author{Takashi Taniguchi}
\affiliation{National Institute for Materials Science, 1-1 Namiki, Tsukuba 305-0044, Japan}

\author{Andrei B. Sushkov}
%\email{}
\affiliation{Center for Nanophysics and Advanced Materials, University of Maryland, College Park, Maryland 20742, USA}

\author{Martin Mittendorff}
%\email{}
\affiliation{Institute for Research in Electronics \& Applied Physics, University of Maryland, College Park, MD 20742, USA}

\author{James Hone}
\affiliation{Department of Mechanical Engineering, Columbia University, New York, New York 10027, USA}

\author{H. Dennis Drew}
%\email{hdrew@umd.edu}
\affiliation{Center for Nanophysics and Advanced Materials, University of Maryland, College Park, Maryland 20742, USA}

\author{Michael S. Fuhrer}
%\email{michael.fuhrer@monash.edu }
\affiliation{Center for Nanophysics and Advanced Materials, University of Maryland, College Park, Maryland 20742, USA}
\affiliation{School of Physics and Astronomy, Monash University, 3800 Victoria, Australia}

\author{Thomas E. Murphy}
\email{tem@umd.edu}
\affiliation{Institute for Research in Electronics \& Applied Physics, University of Maryland, College Park, MD 20742, USA}

\begin{abstract}

Hot electron effects in graphene are significant because of graphene's small electronic heat capacity and weak electron-phonon coupling, yet the dynamics and cooling mechanisms of hot electrons in graphene are not completely understood.  We describe a novel photocurrent spectroscopy method that uses the mixing of continuous-wave lasers in a graphene photothermal detector to measure the frequency dependence and nonlinearity of hot-electron cooling in graphene as a function of the carrier concentration and temperature.  The method offers unparalleled sensitivity to the nonlinearity, and probes the ultrafast cooling of hot carriers with an optical fluence that is orders of magnitude smaller than in conventional time-domain methods, allowing for accurate characterization of electron-phonon cooling near charge neutrality.  Our measurements reveal that near the charge neutral-point the nonlinear power dependence of the electron cooling is dominated by disorder-assisted collisions, while at higher carrier concentrations conventional momentum-conserving cooling prevails in the nonlinear dependence.  The relative contribution of these competing mechanisms can be electrostatically tuned through the application of a gate voltage -- an effect that is unique to graphene.

\end{abstract}%

\maketitle

When graphene absorbs electromagnetic radiation, its electrons heat up and produce a measurable thermoelectric response, even at room temperature.  Because of graphene's gapless dispersion relation, small electronic heat capacity, and anomalously weak electron-phonon coupling, this photothermal detection mechanism is broadband (from dc to visible), highly sensitive, and fast \cite{Zuev2009,Xu2009,Gabor2011,Koppens2014,Cai2014}.  The speed, temperature dependence, and power dependence of these detectors depend critically upon how fast and by what mechanisms the hot carriers relax \cite{Song2012,Graham2013,Ma2014}.  Two primary cooling mechanisms have been identified:  supercollision cooling, in which disorder-assisted scattering allows for non-momentum-conserving transitions, and conventional momentum-conserving electron-phonon cooling \cite{Bistritzer2009,Tse2009,Song2012,Betz2012,Betz2013,Graham2013,Shiue2015, McKitterick2016}. Evidence for conventional momentum-conserving cooling (which is linear with temperature) has been observed only at low temperatures in high-quality graphene\cite{Ma2014}.  In experimental measurements, the cooling process is inferred from how the photothermal response depends on temperature, power, or time for either pulsed or continuous-wave illumination.  Time-domain methods that are used to study thermal relaxation dynamics typically employ intense optical pulses, which significantly disturb the electron temperature, and can in some cases excite higher energy optical phonons in addition to acoustic phonons \cite{Wang2010,Winnerl2011,Tielrooij2013,Graham2013}.  Moreover, as we show here, the factors that govern the power dependence of the photothermal response can be different from those that determine the cooling rate.  It has been shown that, uniquely in graphene, the relative strength of the two competing cooling channels can be controlled by the carrier concentration \cite{Song2012,Betz2013,Ma2014}.

Here we employ a new nonlinear photomixing method to simultaneously quantify the nonlinearity in the photoresponse and the carrier-density dependence of electron cooling in graphene.  This method easily distinguishes between sublinear and superlinear power dependence, which indicate supercollision cooling and conventional cooling, respectively.  Our measurements show that while supercollision cooling dominates the nonlinear response near the charge neutral point, at higher carrier densities, conventional cooling is the dominant contribution to the nonlinearity.  Furthermore, we show that when two detuned near-IR lasers co-illuminate the graphene, the resulting dc photovoltage depends upon their heterodyne difference frequency.   This enables the direct measurement of the electron cooling rate in the frequency domain with orders of magnitude weaker optical excitation (smaller temperature rise) than traditional time-domain methods, by simply tuning the wavelength of one of the continuous-wave lasers.

\begin{figure}[htbp]
  \centering
  \includegraphics{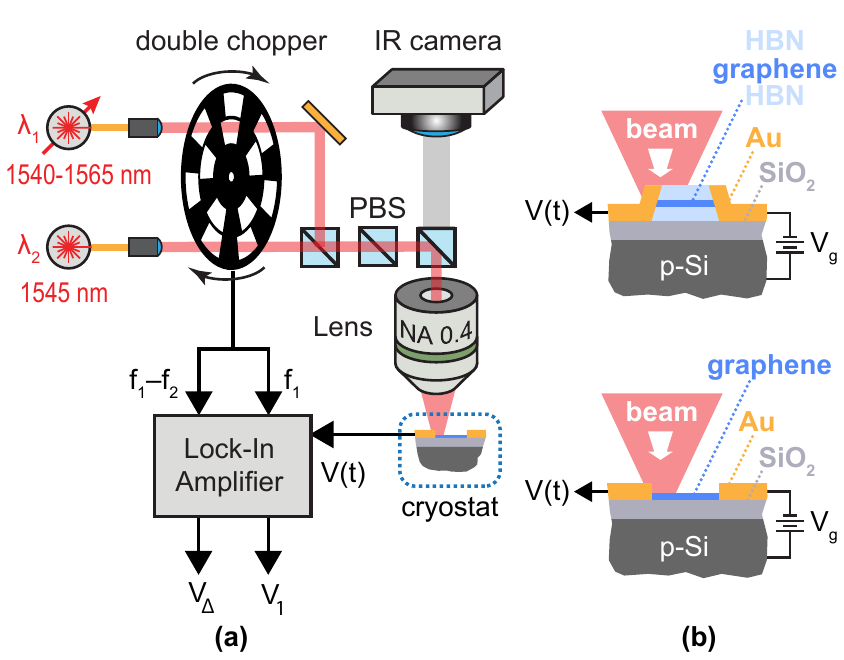}
  \caption{Diagram of heterodyne photomixing experiment. (a) Two near-IR continuous-wave beams, one with a tunable wavelength, are modulated at two different frequencies $f_1$ and $f_2$, then overlapped and focused down to the graphene photodetector. The photovoltages produced by laser 1 ($V_1$) and the mixing of two beams ($V_\Delta$) are detected at the modulation frequencies $f_1$ and $f_1-f_2$ respectively. (b) Diagram of the HBN-encapsulated graphene (top) and exfoliated graphene (bottom) photodetector devices. The optical beams are illuminated close to one of the metal contacts, and the carrier density of graphene is altered by applying an electrostatic voltage ($V_g$) between the doped silicon substrate and graphene.}
  \label{fig:1}
\end{figure}

Figure~\ref{fig:1}(a) depicts the heterodyne photomixing setup used here to characterize the photothermal response of graphene.  Two fiber-coupled continuous-wave near-IR lasers, one wavelength tunable ($\lambda_1 = $ 1540--1565 nm) and one at fixed-wavelength ($\lambda_2=$ 1545 nm), were amplified, spatially combined, polarized, and focused using an aspheric lens to a 3 $\mu$m spot on the graphene channel.  The position of the focused beam was chosen to maximize the photovoltage, which occurs when the beam is focused close to one of the contacts \cite{Lee2008,Xia2009}.  The combined optical power illuminating the first (second) device was about 6 (2.1) mW, from which we estimate the total absorbed intensity to be $I =$ 850 (300) W/cm$^2$.  The graphene photodetector device was held in a liquid helium cryostat with short working distance optical access to controllably vary the lattice temperature $T_L$ between 10 K and room temperature.  The two lasers were mechanically chopped using a twin-slot (5/7) chopping wheel at frequencies $f_1 = $ 500 Hz and $f_2 = $ 700 Hz.  The photovoltage was synchronously detected at both $f_1$ and the difference $f_2-f_1$, using a dual-reference digital lock-in amplifier (Signal Recovery 7270), which simultaneously records the photovoltages $V_1$ and $V_\Delta$.  The phases of the two lock-in detection channels were calibrated to produce the correct sign, relative to one another.  Measurements were performed as a function of the gate voltage $V_g$, and the optical difference frequency $\Delta\nu = \Omega/2\pi$, which was swept from $-0.6$ to $+2.5$ THz by tuning laser 1.

To better elucidate the role of disorder, we considered two different graphene detectors shown in Fig.~\ref{fig:1}(b):  one using an edge-contacted hexagonal boron nitride (HBN) encapsulated graphene channel\cite{Dean2010,Wang2013}, and a second fabricated from an unencapsulated exfoliated flake\cite{Cai2014}. The Supplemental Material\cite{SupplementaryContinuedReferences} details the fabrication and dc electrical characterization of the devices (Sec. S4).

The electron temperature $T$ in the graphene evolves according to the nonlinear differential equation \cite{Bistritzer2009,Viljas2010,Song2012}
\begin{equation}\label{eq:1}
  \alpha T \frac{dT}{dt} + \beta_1 (T - T_L) + \beta_3 (T^3-T_L^3) = I(t)
\end{equation}
where $T_L$ is the lattice temperature, $\alpha T$ is the specific heat of the graphene carriers, the coefficients $\beta_1$ and $\beta_3$ are the rate coefficients for the conventional and supercollision cooling mechanisms, respectively, and $I(t)$ is the absorbed near-infrared optical intensity.  For the two-laser illumination shown in Fig.~\ref{fig:1}(a), the absorbed intensity is $I(t)= I_1 + I_2 + 2\sqrt{I_1I_2}\cos\Omega t$, where $I_1$ and $I_2$ represent the absorbed intensities of lasers 1 and 2, respectively and $\Omega \equiv 2\pi c(\lambda_2^{-1}-\lambda_1^{-1})$ is their heterodyne difference frequency.

It is assumed that the electrons are in the degenerate regime $(E_F \gg k_BT)$, and that the electron-electron collisions are fast enough ($\tau_{\rm ee}^{-1} \gg \Omega$) to allow the temperature of the electron gas to be well defined \cite{Bistritzer2009,DasSarma2011}. The hot-electron diffusion length is $\xi = (\kappa/\gamma\alpha T)^{1/2} = v_F(\gamma\Gamma)^{-1/2}$, where $\kappa$ is the electronic thermal conductivity, $\Gamma$ is the carrier scattering rate, and the Wiedemann-Franz law was used in the second equality. Even for the encapsulated device considered here, by estimating $\Gamma$ from dc measurements in Fig.~S3, we estimate that 500 nm $< \xi < 1.5$ $\mu$m, which is smaller than the optical beam size employed. We therefore ignore spatial inhomogeneity in $I(t)$ and the thermal diffusion of hot carriers out of the laser beam.

The three model parameters $\alpha$, $\beta_1$, and $\beta_3$ appearing in Eq.~\eqref{eq:1} depend implicitly on the Fermi level $E_F$ (determined by gating) and the disorder mean-free path $l$ (related to the quality of the graphene) as $\alpha=2\pi k_B^2 E_F/(3\hbar^2 v_F^2)$, $\beta_1= V_D^2 E_F^4 k_B/(2\pi \rho \hbar^5 v_F^6)$, $\beta_3=\zeta(3) V_D^2 E_F k_B^3/(\pi^2 \rho \hbar^4 v_F^3 s^2 l)$, where $v_F$ is the Fermi velocity, $\rho$ is the areal mass density, $s$ is the speed of sound in graphene, $\zeta(3)\simeq 1.202$ is the Riemann zeta function, and $V_D$ is the acoustic deformation potential.  We note that the substrate surface polar phonons may also play a role in hot electron cooling in graphene\cite{Low2012,Hwang2013,Principi2016}, and their effect on the photoresponse can be regarded as a linear cooling term ($\beta_1$) in Eq.~\eqref{eq:1}\cite{Freitag2013}. At temperatures far below the Bloch-Gr\"uneisen temperature Eq.~\eqref{eq:1} must be modified to include a cooling term proportional to $T^4$\cite{Viljas2010}.  We estimate that even at the lowest temperatures and carrier concentrations experimentally considered here ($T = 25$ K, $E_F \sim 60$ meV), the measurement temperature matches or exceeds the Bloch-Gr\"uneisen temperature.

The resulting photothermoelectric voltage $V$ produced by the Seebeck effect is then related to the electron temperature by $V = rT(T-T_L)$, where $rT$ is the Seebeck coefficient of graphene\cite{Xu2009,Song2011}.  This nonlinear relationship between temperature and photovoltage could be generalized to include a nonlinearity in the Seebeck coefficient\cite{Duan2016,Hwang2009}, but the temperature dependence and power dependence of the observed nonlinearity indicate that this effect is small in comparison to the nonlinearity in cooling.  Although other photoresponse mechanisms, such as the photoelectric effect\cite{Echtermeyer2014}, might also contribute to the graphene photoresponse, a photothermoelectric model can adequately describe the photoresponse at the graphene-metal interface\cite{Cai2014,Shiue2015,Tielrooij2015}.

Equation~\eqref{eq:1} can be solved using a power series expansion (Supplemental Material\cite{SupplementaryContinuedReferences}, Sec.~S1), and the resulting dc photovoltage is found to be
\begin{equation}
  \label{eq:2}
  \begin{split}
    V(I_1,I_2) =\:&a_1 (I_1+I_2) + a_2 (I_1^2+I_2^2) - a_3 (I_1^3 + I_2^3) \quad\ldots \\
      +\:&2a_2I_1I_2\biggl(1+\frac{\gamma^2}{\Omega^2 + \gamma^2}\biggr) \quad\ldots \\
      -\:&3a_3I_1I_2(I_1+I_2)\biggl(1+\frac{2\gamma^2}{\Omega^2 + \gamma^2}\biggr)
  \end{split}
\end{equation}
where $\gamma \equiv (\beta_1 + 3\beta_3 T_L^2)/\alpha T_L$ is the linearized cooling rate from both mechanisms.  The coefficients $a_1$, $a_2$, and $a_3$ are given by
\begin{equation}\label{eq:3}
  a_1 \equiv \frac{r}{\alpha\gamma},\quad a_2 \equiv \frac{r\beta_1}{(\alpha\gamma T_L)^3},\quad a_3 \equiv \frac{3r\beta_3^2}{T_L^2\alpha^5\gamma^5}
\end{equation}

The final two terms in Eq.~\eqref{eq:2} which contain the factor $I_1I_2$, represent a nonlinear interaction of the two beams, which occurs only when both beams are present.  In order to sensitively detect only these mixing products, we employ a double-modulation configuration in which laser 1 is mechanically chopped at a frequency $f_1$, laser 2 is chopped at $f_2$, and the photovoltage $V$ is synchronously detected using a lock-in amplifier at the chopping difference frequency $\Delta f \equiv f_1-f_2$ (not to be confused with the heterodyne difference frequency).  The resulting photovoltage $V_\Delta \equiv V(I_1,I_2) - V(I_1,0) - V(0,I_2)$ can be positive or negative, depending on the nonlinearity in the photothermal response.  We also simultaneously measure the Fourier component at $f_1$, denoted $V_1 \equiv V(I_1,0)$, which gives the photovoltage produced by laser 1 alone.

By simply comparing the magnitude of the two terms that compose the linearized cooling rate $\gamma$, one can determine a condition for which process makes the largest contribution to the cooling rate.  For nearly all of the experimental cases considered here and reported elsewhere, the cooling rate is largely limited by the supercollision term.  Equation \eqref{eq:3} reveals that despite this, the photoresponse can be either superlinear or sublinear in intensity, depending on the carrier density and graphene quality.  As explained below, neither cooling effect can be ignored when analyzing the nonlinearity of the response.

When the heterodyne frequency exceeds the cooling rate ($\Omega \gg \gamma$), Eq.~\eqref{eq:2} simplifies to $V(I) = a_1 I + a_2 I^2 - a_3 I^3$, where $I \equiv I_1+I_2$ is the total absorbed optical intensity.  The quadratic and cubic terms have opposite sign, and therefore describe superlinear or sublinear dependence on the optical intensity.  From Eq.~\eqref{eq:3}, one sees that the superlinear coefficient is proportional to $\beta_1$, which we associate with momentum-conserving cooling, while the sublinear coefficient is proportional to $\beta_3$, which arises from supercollision cooling.

\begin{figure*}[htbp]
  \centering
  \includegraphics{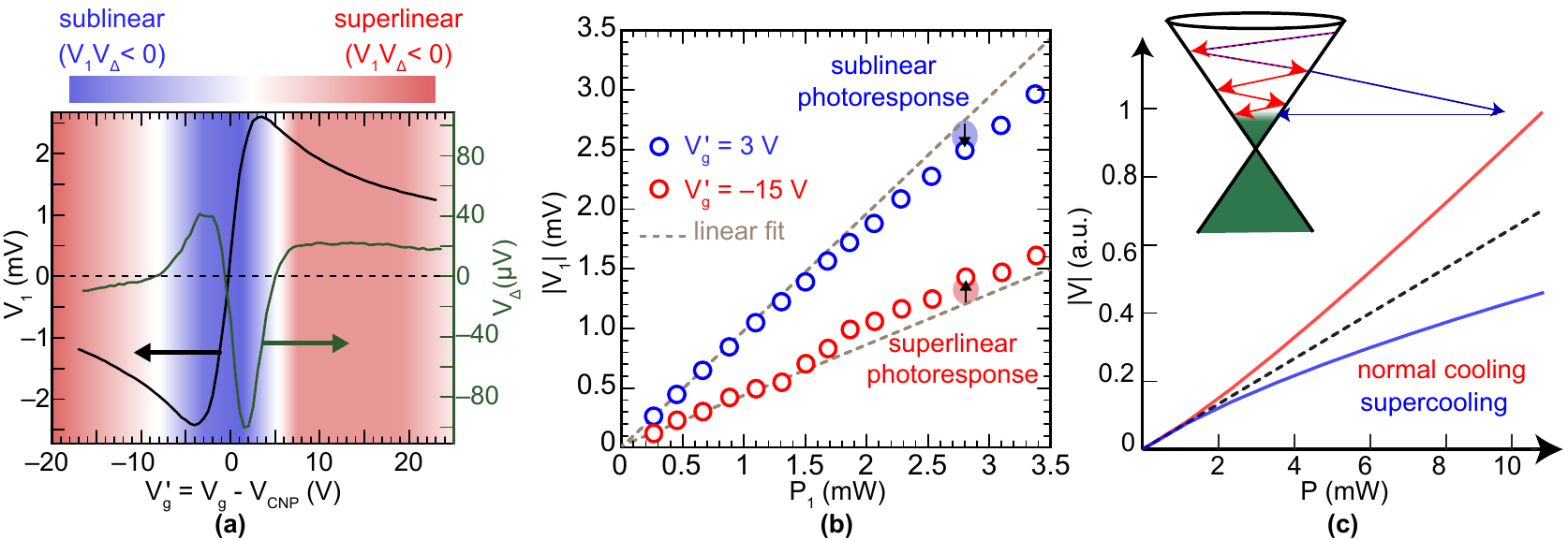}
  \caption{Photovoltage versus carrier concentration and incident power, measured at room temperature. (a) Single-laser photovoltage ($V_1$), and nonlinear photomixing signal ($V_\Delta$) measured versus the applied gate voltage $V_g$.  $V_{\rm CNP}$ denotes the charge neutral point. The red ($V_1V_\Delta>0$) and blue ($V_1V_\Delta<0$) regions indicate the conditions where super- and sublinear power dependence is observed. (b) Measured photovoltage $V_1$ of a single laser versus incident optical power at two different gate voltages, showing sublinear and superlinear behavior.  (c) Calculated photothermoelectric voltage (in arbitrary units) versus input optical power for the case of $\beta_1=0$ (blue) and $\beta_3=0$ (red), illustrating the sublinear and superlinear behavior, respectively. Inset: energy diagram illustrating the two different cooling mechanisms. }
  \label{fig:2}
\end{figure*}

Figure~\ref{fig:2}(a) plots $V_1$ (black) and $V_\Delta$ (green) as a function of the gate voltage for the HBN-encapsulated device.  These measurements were performed with $\Omega/2\pi = $ 2.5 THz, which is much faster than the expected cooling rate at room temperature.  The sign of the photothermal voltage $V_1$ depends on the gate voltage, as expected from the photothermoelectric effect \cite{Xu2009,Gabor2011,Cai2014}.  For carrier densities near the charge neutral point, $V_1$ and $V_\Delta$ have opposite sign (as indicated by the blue shading), revealing a sublinear power dependence, characteristic of supercollision cooling \cite{Graham2013,Shiue2015}.  In this regime the Fermi surface is small, and the allowed phonon energy space for the momentum-conserving collision is strongly constrained, thereby suppressing conventional electron-phonon cooling \cite{Song2012,Betz2013}.  At higher carrier densities, the behavior changes to superlinear (red shading), indicating that conventional cooling becomes stronger and dominates the photothermal nonlinearity.  Figure~\ref{fig:2}(b) plots the single-beam photovoltage as a function of the incident optical power, confirming the sub- and superlinear behavior, respectively.  Figure~\ref{fig:2}(c) illustrates the two cooling mechanisms schematically in $k$ space, along with the predicted sublinear and superlinear power dependence.  The transitions outside of the Dirac cone represent supercollision cooling, in which the spatial disorder in the graphene compensates for the electron-phonon momentum mismatch.

The threshold between these two nonlinear regimes can be approximated by equating the opposing terms in $V_\Delta$, which gives
\begin{equation}\label{eq:4}
  2a_2 \gtrless 3a_3 I
\end{equation}
where the upper and lower inequalities describe the conditions under which conventional cooling or supercollision cooling prevails in the nonlinear response, respectively.  The relative importance of the two competing cooling channels depends on temperature, intensity, the carrier concentration (Fermi level), and indirectly on the material quality, which is related to the disorder mean-free path $l$. Even though the linearized cooling rate $\gamma$ is limited by supercollision cooling, both effects are evident in the nonlinear response reported here.

In the Supplemental Material (Ref.~\cite{SupplementaryContinuedReferences}, Sec.~S2), we present the results of a similar measurement performed on lower-mobility exfoliated graphene on SiO$_2$. Similar to the HBN-encapsulated device, we observe an expected transition from supercollision cooling to conventional cooling. The transition happens around $E_F=80$ meV, and we use Eq.~\eqref{eq:4} to determine the ratio of the two rate coefficients, $\beta_1/\beta_3 = 5300$ $\rm{K}^2$.  At room temperature, the supercollision contribution to the cooling rate $\gamma$ is nearly 50$\times$ larger than the contribution from conventional cooling.  Despite this, both effects have a non-negligible role in the nonlinearity of the photoresponse, and their relative significance depends on the carrier density.

When the heterodyne difference frequency $\Omega$ is comparable to or smaller than the cooling rate $\gamma$, the electron temperature can follow the interferometric beating of the two lasers, which produces a larger photothermal voltage than when the lasers are widely detuned.  The final two terms in Eq.~\eqref{eq:2} reveal that the nonlinear mixing signal $V_\Delta$ exhibits a Lorentzian dependence on the heterodyne difference frequency $\Omega$, with a spectral width that is proportional to the cooling rate $\gamma$.  As before, the double-chopping configuration allows for sensitive detection of this heterodyne photomixing signal.

\begin{figure}[htbp]
  \centering
  \includegraphics[]{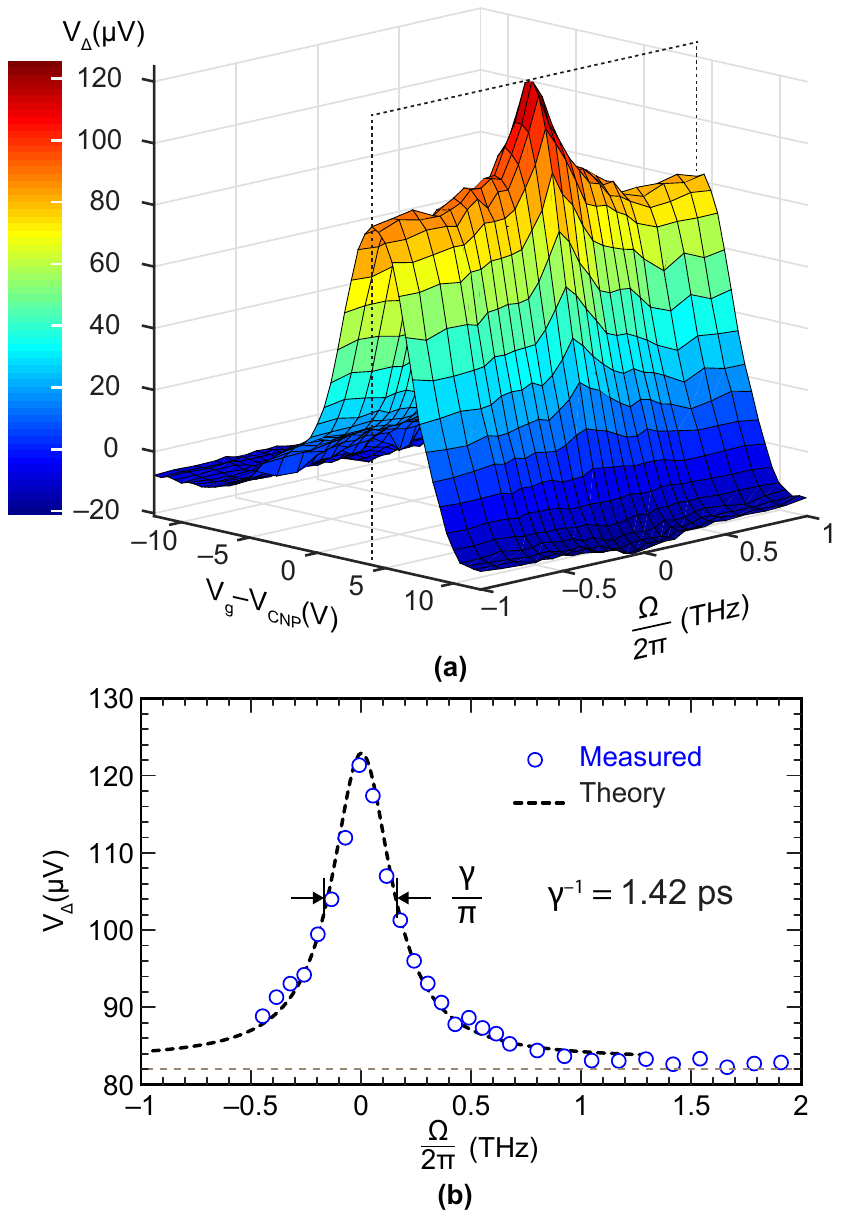}
  \caption{Heterodyne photomixing response for the nonencapsulated device, measured at room temperature. (a) The nonlinear photomixing signal ($V_{\Delta}$) as a function of the difference frequency ($\Omega$) and gate voltage ($V_g$). At each gate voltage, $V_\Delta$ exhibits a Lorentzian-shaped dependence on the heterodyne difference frequency $\Omega$.  (b) $V_\Delta$ vs $\Omega$ for $V_g = 4$ V. The dashed curve is the theoretical calculation of the photomixing voltage based on a photothermoelectric effect. From the Lorentzian fit, the hot electron cooling time is estimated to be $\gamma^{-1} = 1.42$ ps.}
  \label{fig:3}
\end{figure}

Figure~\ref{fig:3}(a) plots the measured photovoltage $V_\Delta$ as a function of the gate voltage and heterodyne difference frequency, for the nonencapsulated graphene detector.  In addition to the expected gate-voltage dependence discussed previously, the photoresponse exhibits a distinct spectral peak around $\Omega = 0$.  Figure~\ref{fig:3}(b) shows the photomixing spectrum at a fixed gate voltage, along with the best-fit Lorentzian curve.  From the linewidth, we estimate a cooling time of $\gamma^{-1} = $1.42 ps, which is consistent with the time-domain pulse coincidence measurements \cite{Sun2012,Cai2014} reported for similar devices.

\begin{figure}[htbp]
  \centering
  \includegraphics[]{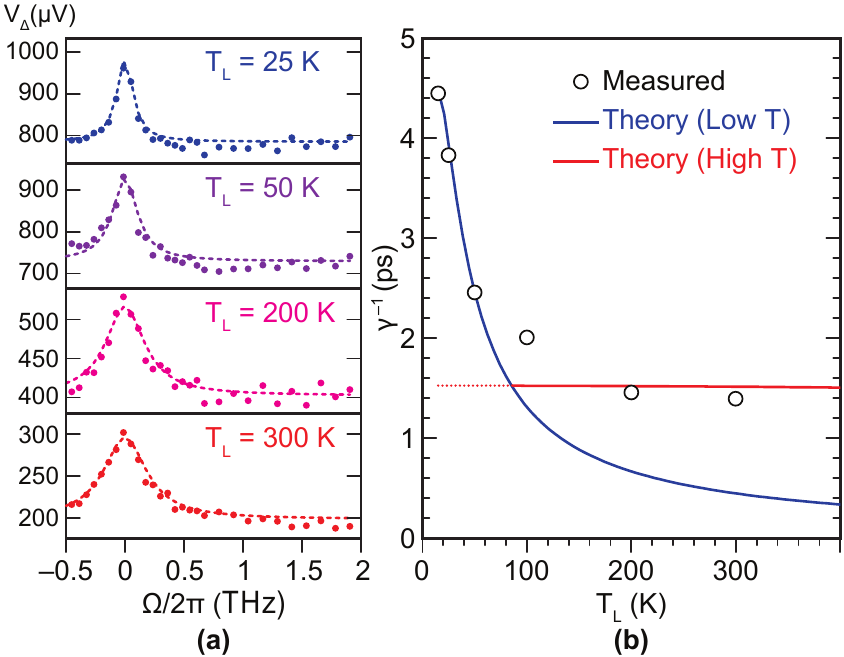}
  \caption{Temperature dependence of the cooling rate.  (a) The two-beam mixing signal as a function of the difference frequency measured close to the charge neutral point (sublinear power dependence regime) at different temperatures for the exfoliated sample on Si$\rm{O}_2$. The dashed curves are Lorentzian fits. (b) The black circles are the extracted hot electron cooling time constant $\tau \equiv \gamma^{-1}$ from the Lorentzian fits in panel (a). The blue (red) curve is the theory fit for low (high) temperature.}
  \label{fig:4}
\end{figure}

In order to confirm the thermal model for the photomixing, we repeated the heterodyne spectral measurements at temperatures from room temperature down to 25 K for the exfoliated sample on Si$\rm{O}_2$ near the charge neutral point.  As shown in Fig.~\ref{fig:4}(a), in all cases the photomixing signal exhibits a Lorentzian spectral dependence, but with a spectral width that decreases with temperature, as expected.  The solid blue curve in Fig.~\ref{fig:4}(b) shows a fit to the linearized cooling rate, based on the model presented here.  For the data points above $T_L = 80 K$, and for the conditions at the charge neutral point, the assumption of $E_F \gg k_BT$ (degenerate regime) is no longer valid, which requires a modification of the cooling rate (Supplemental Material\cite{SupplementaryContinuedReferences}, Sec.~S3).  We therefore excluded these points when fitting the blue curve.  However, when the parameters from the low-temperature fit were incorporated into the modified thermal model, it correctly predicts the observed high-temperature asymptotic behavior, indicated by the red curve, with no additional free parameters.

This nonlinear heterodyne photovoltage spectroscopy method has two important advantages over the traditional time-domain measurement using pulse coincidence \cite{Sun2012,Graham2013,Cai2014}: (i) the frequency range and resolution is limited only by the tuning range and resolution of the laser, while in time-domain measurements the response is limited by the optical pulse width and repetition period; (ii) continuous-wave illumination produces a far smaller thermal stimulus to the graphene electrons than intense ultrafast pulses, thereby allowing the measurement of the temporal dynamics and nonlinearity of photodetection under low photothermal excitation for which the electron temperature is near the lattice temperature.

The model and measurements described here show that there are two competing cooling channels for hot electrons in graphene, and Eq.~\eqref{eq:4} describes the relative importance of each in the nonlinear response.  In time-domain experiments reported elsewhere, the instantaneous absorbed intensity is orders of magnitude higher that the continuous-wave illumination considered here, in which case Eq.~\eqref{eq:4} predicts that supercollision cooling is the dominant contribution to the nonlinearity at all practically attainable doping concentrations.  Moreover nonencapsulated graphene samples have a much smaller disorder mean-free path $l$, which further contributes to the relative importance of supercollision cooling over conventional momentum-conserving cooling.  In these cases, the photothermal response is often adequately described by supercollision cooling alone, for a wide range of carrier densities and temperatures \cite{MWGraham2013,Shiue2015,Jadidi2016}.  For continuous-wave measurements on encapsulated devices, Eq.~\eqref{eq:4} also predicts that at sufficiently low temperatures, conventional cooling will prevail, consistent with temperature-dependent measurements reported recently\cite{Ma2014}.

We show that nonlinearity in the photothermoelectric effect causes photomixing when graphene is illuminated by near-infrared beams, and we describe a new heterodyne spectroscopy method that accurately measures this nonlinearity in the frequency domain. Exceedingly small nonlinearities in the photoresponse can be probed using continuous-wave illumination, which accurately elucidates the physical mechanisms behind the nonlinearity and cooling.  In particular, the measurements reveal the role that disorder plays in the cooling of hot electrons, and the interplay of different cooling channels at different carrier concentrations.  The method permits the direct measurement of the cooling rate in graphene using swept laser spectroscopy, which offers several advantages over pump-probe or pulse-coincidence measurements. The work also implies that nonlinear photomixing in graphene is very promising for the development of new optical and THz photomixing devices.

This work was sponsored by the U.S. ONR (N000141310865) and the U.S. NSF (ECCS 1309750).  C.T. was supported by a DOD-AFOSR, NDSEG fellowship under Contract No. FA9550-11-C-0028, 32 CFR 168a. C.T. and J.H. acknowledge financial support from the Nanoelectronics Research Initiative (NRI) through the Institute for Nanoelectronics Discovery and Exploration (INDEX).  K.W. and T.T. acknowledge support from the Elemental Strategy Initiative conducted by the MEXT, Japan and a Grant-in-Aid for Scientific Research on Innovative Areas ``Science of Atomic Layers'' from JSPS. M.S.F. was supported in part by an ARC Laureate Fellowship.

%\bibliography{Jadidi-Mixing-2016}

\begin{thebibliography}{35}%
\makeatletter
\providecommand \@ifxundefined [1]{%
 \@ifx{#1\undefined}
}%
\providecommand \@ifnum [1]{%
 \ifnum #1\expandafter \@firstoftwo
 \else \expandafter \@secondoftwo
 \fi
}%
\providecommand \@ifx [1]{%
 \ifx #1\expandafter \@firstoftwo
 \else \expandafter \@secondoftwo
 \fi
}%
\providecommand \natexlab [1]{#1}%
\providecommand \enquote  [1]{``#1''}%
\providecommand \bibnamefont  [1]{#1}%
\providecommand \bibfnamefont [1]{#1}%
\providecommand \citenamefont [1]{#1}%
\providecommand \href@noop [0]{\@secondoftwo}%
\providecommand \href [0]{\begingroup \@sanitize@url \@href}%
\providecommand \@href[1]{\@@startlink{#1}\@@href}%
\providecommand \@@href[1]{\endgroup#1\@@endlink}%
\providecommand \@sanitize@url [0]{\catcode `\\12\catcode `\$12\catcode
  `\&12\catcode `\#12\catcode `\^12\catcode `\_12\catcode `\%12\relax}%
\providecommand \@@startlink[1]{}%
\providecommand \@@endlink[0]{}%
\providecommand \url  [0]{\begingroup\@sanitize@url \@url }%
\providecommand \@url [1]{\endgroup\@href {#1}{\urlprefix }}%
\providecommand \urlprefix  [0]{URL }%
\providecommand \Eprint [0]{\href }%
\providecommand \doibase [0]{http://dx.doi.org/}%
\providecommand \selectlanguage [0]{\@gobble}%
\providecommand \bibinfo  [0]{\@secondoftwo}%
\providecommand \bibfield  [0]{\@secondoftwo}%
\providecommand \translation [1]{[#1]}%
\providecommand \BibitemOpen [0]{}%
\providecommand \bibitemStop [0]{}%
\providecommand \bibitemNoStop [0]{.\EOS\space}%
\providecommand \EOS [0]{\spacefactor3000\relax}%
\providecommand \BibitemShut  [1]{\csname bibitem#1\endcsname}%
\let\auto@bib@innerbib\@empty
%</preamble>
\bibitem [{\citenamefont {Zuev}\ \emph {et~al.}(2009)\citenamefont {Zuev},
  \citenamefont {Chang},\ and\ \citenamefont {Kim}}]{Zuev2009}%
  \BibitemOpen
  \bibfield  {author} {\bibinfo {author} {\bibfnamefont {Y.M.}\ \bibnamefont
  {Zuev}}, \bibinfo {author} {\bibfnamefont {W.}~\bibnamefont {Chang}}, \ and\
  \bibinfo {author} {\bibfnamefont {P.}~\bibnamefont {Kim}},\ }\bibfield
  {title} {\enquote {\bibinfo {title} {{Thermoelectric and
  magnetothermoelectric transport measurements of graphene.}}}\ }\href@noop {}
  {\bibfield  {journal} {\bibinfo  {journal} {Phys. Rev. Lett.}\ }\textbf
  {\bibinfo {volume} {102}},\ \bibinfo {pages} {096807} (\bibinfo {year}
  {2009})}\BibitemShut {NoStop}%
\bibitem [{\citenamefont {Xu}\ \emph {et~al.}(2009)\citenamefont {Xu},
  \citenamefont {Gabor}, \citenamefont {Alden}, \citenamefont {van~der Zande},\
  and\ \citenamefont {McEuen}}]{Xu2009}%
  \BibitemOpen
  \bibfield  {author} {\bibinfo {author} {\bibfnamefont {X.}~\bibnamefont
  {Xu}}, \bibinfo {author} {\bibfnamefont {N.M.}\ \bibnamefont {Gabor}},
  \bibinfo {author} {\bibfnamefont {J.S.}\ \bibnamefont {Alden}}, \bibinfo
  {author} {\bibfnamefont {A.M.}\ \bibnamefont {van~der Zande}}, \ and\
  \bibinfo {author} {\bibfnamefont {P.L.}\ \bibnamefont {McEuen}},\ }\bibfield
  {title} {\enquote {\bibinfo {title} {{Photo-Thermoelectric Effect at a
  Graphene Interface Junction}},}\ }\href@noop {} {\bibfield  {journal}
  {\bibinfo  {journal} {Nano Lett.}\ }\textbf {\bibinfo {volume} {10}},\
  \bibinfo {pages} {562} (\bibinfo {year} {2010})}\BibitemShut {NoStop}%
\bibitem [{\citenamefont {Gabor}\ \emph {et~al.}(2011)\citenamefont {Gabor},
  \citenamefont {Song}, \citenamefont {Ma}, \citenamefont {Nair}, \citenamefont
  {Taychatanapat}, \citenamefont {Watanabe}, \citenamefont {Taniguchi},
  \citenamefont {Levitov},\ and\ \citenamefont {Jarillo-Herrero}}]{Gabor2011}%
  \BibitemOpen
  \bibfield  {author} {\bibinfo {author} {\bibfnamefont {N.M.}\ \bibnamefont
  {Gabor}}, \bibinfo {author} {\bibfnamefont {J.C.}\ \bibnamefont {Song}},
  \bibinfo {author} {\bibfnamefont {Q.}~\bibnamefont {Ma}}, \bibinfo {author}
  {\bibfnamefont {N.L.}\ \bibnamefont {Nair}}, \bibinfo {author} {\bibfnamefont
  {T.}~\bibnamefont {Taychatanapat}}, \bibinfo {author} {\bibfnamefont
  {K.}~\bibnamefont {Watanabe}}, \bibinfo {author} {\bibfnamefont
  {T.}~\bibnamefont {Taniguchi}}, \bibinfo {author} {\bibfnamefont {L.S.}\
  \bibnamefont {Levitov}}, \ and\ \bibinfo {author} {\bibfnamefont
  {P.}~\bibnamefont {Jarillo-Herrero}},\ }\bibfield  {title} {\enquote
  {\bibinfo {title} {{Hot carrier–-assisted intrinsic photoresponse in
  graphene}},}\ }\href@noop {} {\bibfield  {journal} {\bibinfo  {journal}
  {Science}\ }\textbf {\bibinfo {volume} {334}},\ \bibinfo {pages} {648}
  (\bibinfo {year} {2011})}\BibitemShut {NoStop}%
\bibitem [{\citenamefont {Koppens}\ \emph {et~al.}(2014)\citenamefont
  {Koppens}, \citenamefont {Mueller}, \citenamefont {Avouris}, \citenamefont
  {Ferrari}, \citenamefont {Vitiello},\ and\ \citenamefont
  {Polini}}]{Koppens2014}%
  \BibitemOpen
  \bibfield  {author} {\bibinfo {author} {\bibfnamefont {F.H.L.}\ \bibnamefont
  {Koppens}}, \bibinfo {author} {\bibfnamefont {T.}~\bibnamefont {Mueller}},
  \bibinfo {author} {\bibfnamefont {P.}~\bibnamefont {Avouris}}, \bibinfo
  {author} {\bibfnamefont {A.C.}\ \bibnamefont {Ferrari}}, \bibinfo {author}
  {\bibfnamefont {M.S.}\ \bibnamefont {Vitiello}}, \ and\ \bibinfo {author}
  {\bibfnamefont {M.}~\bibnamefont {Polini}},\ }\bibfield  {title} {\enquote
  {\bibinfo {title} {{Photodetectors based on graphene, other two-dimensional
  materials and hybrid systems}},}\ }\href@noop {} {\bibfield  {journal}
  {\bibinfo  {journal} {Nat. Nanotechnol.}\ }\textbf {\bibinfo {volume} {9}},\
  \bibinfo {pages} {780} (\bibinfo {year} {2014})}\BibitemShut {NoStop}%
\bibitem [{\citenamefont {Cai}\ \emph {et~al.}(2014)\citenamefont {Cai},
  \citenamefont {Sushkov}, \citenamefont {Suess}, \citenamefont {Jadidi},
  \citenamefont {Jenkins}, \citenamefont {Nyakiti}, \citenamefont {Myers-Ward},
  \citenamefont {Li}, \citenamefont {Yan}, \citenamefont {Gaskill},
  \citenamefont {Murphy}, \citenamefont {Drew},\ and\ \citenamefont
  {Fuhrer}}]{Cai2014}%
  \BibitemOpen
  \bibfield  {author} {\bibinfo {author} {\bibfnamefont {X.}~\bibnamefont
  {Cai}}, \bibinfo {author} {\bibfnamefont {A.~B.}\ \bibnamefont {Sushkov}},
  \bibinfo {author} {\bibfnamefont {R.~J.}\ \bibnamefont {Suess}}, \bibinfo
  {author} {\bibfnamefont {M.~M.}\ \bibnamefont {Jadidi}}, \bibinfo {author}
  {\bibfnamefont {G.~S.}\ \bibnamefont {Jenkins}}, \bibinfo {author}
  {\bibfnamefont {L.~O.}\ \bibnamefont {Nyakiti}}, \bibinfo {author}
  {\bibfnamefont {R.~L.}\ \bibnamefont {Myers-Ward}}, \bibinfo {author}
  {\bibfnamefont {S.}~\bibnamefont {Li}}, \bibinfo {author} {\bibfnamefont
  {J.}~\bibnamefont {Yan}}, \bibinfo {author} {\bibfnamefont {D.~K.}\
  \bibnamefont {Gaskill}}, \bibinfo {author} {\bibfnamefont {T.~E.}\
  \bibnamefont {Murphy}}, \bibinfo {author} {\bibfnamefont {H.~D.}\
  \bibnamefont {Drew}}, \ and\ \bibinfo {author} {\bibfnamefont {M.~S.}\
  \bibnamefont {Fuhrer}},\ }\bibfield  {title} {\enquote {\bibinfo {title}
  {{Sensitive room-temperature terahertz detection via the photothermoelectric
  effect in graphene}},}\ }\href {\doibase 10.1038/nnano.2014.182} {\bibfield
  {journal} {\bibinfo  {journal} {Nat. Nanotechnol.}\ }\textbf {\bibinfo
  {volume} {9}},\ \bibinfo {pages} {814} (\bibinfo {year}
  {2014})}\BibitemShut {NoStop}%
\bibitem [{\citenamefont {Song}\ \emph {et~al.}(2012)\citenamefont {Song},
  \citenamefont {Reizer},\ and\ \citenamefont {Levitov}}]{Song2012}%
  \BibitemOpen
  \bibfield  {author} {\bibinfo {author} {\bibfnamefont {J.~C.~W.}\
  \bibnamefont {Song}}, \bibinfo {author} {\bibfnamefont {M.~Y.}\ \bibnamefont
  {Reizer}}, \ and\ \bibinfo {author} {\bibfnamefont {L.~S.}\ \bibnamefont
  {Levitov}},\ }\bibfield  {title} {\enquote {\bibinfo {title}
  {{Disorder-Assisted Electron-Phonon Scattering and Cooling Pathways in
  Graphene}},}\ }\href {\doibase 10.1103/PhysRevLett.109.106602} {\bibfield
  {journal} {\bibinfo  {journal} {Phys. Rev. Lett.}\ }\textbf {\bibinfo
  {volume} {109}},\ \bibinfo {pages} {106602} (\bibinfo {year}
  {2012})}\BibitemShut {NoStop}%
\bibitem [{\citenamefont {Graham}\ \emph
  {et~al.}(2013{\natexlab{a}})\citenamefont {Graham}, \citenamefont {Shi},
  \citenamefont {Ralph}, \citenamefont {Park},\ and\ \citenamefont
  {McEuen}}]{Graham2013}%
  \BibitemOpen
  \bibfield  {author} {\bibinfo {author} {\bibfnamefont {M.W.}\ \bibnamefont
  {Graham}}, \bibinfo {author} {\bibfnamefont {S.F.}\ \bibnamefont {Shi}},
  \bibinfo {author} {\bibfnamefont {D.C.}\ \bibnamefont {Ralph}}, \bibinfo
  {author} {\bibfnamefont {J.}~\bibnamefont {Park}}, \ and\ \bibinfo {author}
  {\bibfnamefont {P.L.}\ \bibnamefont {McEuen}},\ }\bibfield  {title} {\enquote
  {\bibinfo {title} {{Photocurrent measurements of supercollision cooling in
  graphene}},}\ }\href@noop {} {\bibfield  {journal} {\bibinfo  {journal} {Nat.
  Phys.}\ }\textbf {\bibinfo {volume} {9}},\ \bibinfo {pages} {103}
  (\bibinfo {year} {2013}{\natexlab{a}})}\BibitemShut {NoStop}%
\bibitem [{\citenamefont {Ma}\ \emph {et~al.}(2014)\citenamefont {Ma},
  \citenamefont {Gabor}, \citenamefont {Andersen}, \citenamefont {Nair},
  \citenamefont {Watanabe}, \citenamefont {Taniguchi},\ and\ \citenamefont
  {Jarillo-Herrero}}]{Ma2014}%
  \BibitemOpen
  \bibfield  {author} {\bibinfo {author} {\bibfnamefont {Q.}~\bibnamefont
  {Ma}}, \bibinfo {author} {\bibfnamefont {N.M.}\ \bibnamefont {Gabor}},
  \bibinfo {author} {\bibfnamefont {T.I.}\ \bibnamefont {Andersen}}, \bibinfo
  {author} {\bibfnamefont {N.L.}\ \bibnamefont {Nair}}, \bibinfo {author}
  {\bibfnamefont {K.}~\bibnamefont {Watanabe}}, \bibinfo {author}
  {\bibfnamefont {T.}~\bibnamefont {Taniguchi}}, \ and\ \bibinfo {author}
  {\bibfnamefont {P.}~\bibnamefont {Jarillo-Herrero}},\ }\bibfield  {title}
  {\enquote {\bibinfo {title} {{Competing channels for hot-electron cooling in
  graphene.}}}\ }\href@noop {} {\bibfield  {journal} {\bibinfo  {journal}
  {Phys. Rev. Lett.}\ }\textbf {\bibinfo {volume} {112}},\ \bibinfo {pages}
  {247401} (\bibinfo {year} {2014})}\BibitemShut {NoStop}%
\bibitem [{\citenamefont {Bistritzer}\ and\ \citenamefont
  {MacDonald}(2009)}]{Bistritzer2009}%
  \BibitemOpen
  \bibfield  {author} {\bibinfo {author} {\bibfnamefont {R.}~\bibnamefont
  {Bistritzer}}\ and\ \bibinfo {author} {\bibfnamefont {A.~H.}\ \bibnamefont
  {MacDonald}},\ }\bibfield  {title} {\enquote {\bibinfo {title} {{Electronic
  Cooling in Graphene}},}\ }\href {\doibase 10.1103/PhysRevLett.102.206410}
  {\bibfield  {journal} {\bibinfo  {journal} {Phys. Rev. Lett.}\ }\textbf
  {\bibinfo {volume} {102}},\ \bibinfo {pages} {206410} (\bibinfo {year}
  {2009})}\BibitemShut {NoStop}%
\bibitem [{\citenamefont {Tse}\ and\ \citenamefont
  {Das~Sarma}(2009)}]{Tse2009}%
  \BibitemOpen
  \bibfield  {author} {\bibinfo {author} {\bibfnamefont {W.-K.}\ \bibnamefont
  {Tse}}\ and\ \bibinfo {author} {\bibfnamefont {S.}~\bibnamefont
  {Das~Sarma}},\ }\bibfield  {title} {\enquote {\bibinfo {title} {{Energy
  relaxation of hot Dirac fermions in graphene}},}\ }\href {\doibase
  10.1103/PhysRevB.79.235406} {\bibfield  {journal} {\bibinfo  {journal} {Phys.
  Rev. B}\ }\textbf {\bibinfo {volume} {79}},\ \bibinfo {pages} {235406}
  (\bibinfo {year} {2009})}\BibitemShut {NoStop}%
\bibitem [{\citenamefont {Betz}\ \emph {et~al.}(2012)\citenamefont {Betz},
  \citenamefont {Vialla}, \citenamefont {Brunel}, \citenamefont {Voisin},
  \citenamefont {Picher}, \citenamefont {Cavanna}, \citenamefont {Madouri},
  \citenamefont {F\`eve}, \citenamefont {Berroir}, \citenamefont
  {Pla\c{c}ais},\ and\ \citenamefont {Pallecchi}}]{Betz2012}%
  \BibitemOpen
  \bibfield  {author} {\bibinfo {author} {\bibfnamefont {A.~C.}\ \bibnamefont
  {Betz}}, \bibinfo {author} {\bibfnamefont {F.}~\bibnamefont {Vialla}},
  \bibinfo {author} {\bibfnamefont {D.}~\bibnamefont {Brunel}}, \bibinfo
  {author} {\bibfnamefont {C.}~\bibnamefont {Voisin}}, \bibinfo {author}
  {\bibfnamefont {M.}~\bibnamefont {Picher}}, \bibinfo {author} {\bibfnamefont
  {A.}~\bibnamefont {Cavanna}}, \bibinfo {author} {\bibfnamefont
  {A.}~\bibnamefont {Madouri}}, \bibinfo {author} {\bibfnamefont
  {G.}~\bibnamefont {F\`eve}}, \bibinfo {author} {\bibfnamefont {J.-M.}\
  \bibnamefont {Berroir}}, \bibinfo {author} {\bibfnamefont {B.}~\bibnamefont
  {Pla\c{c}ais}}, \ and\ \bibinfo {author} {\bibfnamefont {E.}~\bibnamefont
  {Pallecchi}},\ }\bibfield  {title} {\enquote {\bibinfo {title} {{Hot Electron
  Cooling by Acoustic Phonons in Graphene}},}\ }\href {\doibase
  10.1103/PhysRevLett.109.056805} {\bibfield  {journal} {\bibinfo  {journal}
  {Phys. Rev. Lett.}\ }\textbf {\bibinfo {volume} {109}},\ \bibinfo {pages}
  {056805} (\bibinfo {year} {2012})}\BibitemShut {NoStop}%
\bibitem [{\citenamefont {Betz}\ \emph {et~al.}(2013)\citenamefont {Betz},
  \citenamefont {Vialla}, \citenamefont {Brunel}, \citenamefont {Voisin},
  \citenamefont {Picher}, \citenamefont {Cavanna}, \citenamefont {Madouri},
  \citenamefont {F\`eve}, \citenamefont {Berroir}, \citenamefont
  {Pla\c{c}ais},\ and\ \citenamefont {Pallecchi}}]{Betz2013}%
  \BibitemOpen
  \bibfield  {author} {\bibinfo {author} {\bibfnamefont {A.~C.}\ \bibnamefont
  {Betz}}, \bibinfo {author} {\bibfnamefont {F.}~\bibnamefont {Vialla}},
  \bibinfo {author} {\bibfnamefont {D.}~\bibnamefont {Brunel}}, \bibinfo
  {author} {\bibfnamefont {C.}~\bibnamefont {Voisin}}, \bibinfo {author}
  {\bibfnamefont {M.}~\bibnamefont {Picher}}, \bibinfo {author} {\bibfnamefont
  {A.}~\bibnamefont {Cavanna}}, \bibinfo {author} {\bibfnamefont
  {A.}~\bibnamefont {Madouri}}, \bibinfo {author} {\bibfnamefont
  {G.}~\bibnamefont {F\`eve}}, \bibinfo {author} {\bibfnamefont {J.-M.}\
  \bibnamefont {Berroir}}, \bibinfo {author} {\bibfnamefont {B.}~\bibnamefont
  {Pla\c{c}ais}}, \ and\ \bibinfo {author} {\bibfnamefont {E.}~\bibnamefont
  {Pallecchi}},\ }\bibfield  {title} {\enquote {\bibinfo {title}
  {{Supercollision cooling in undoped graphene}},}\ }\href@noop {} {\bibfield
  {journal} {\bibinfo  {journal} {Nat. Phys.}\ }\textbf {\bibinfo {volume}
  {9}},\ \bibinfo {pages} {109} (\bibinfo {year} {2013})}\BibitemShut
  {NoStop}%
\bibitem [{\citenamefont {Shiue}\ \emph {et~al.}(2015)\citenamefont {Shiue},
  \citenamefont {Gao}, \citenamefont {Wang}, \citenamefont {Peng},
  \citenamefont {Robertson}, \citenamefont {Efetov}, \citenamefont {Assefa},
  \citenamefont {Koppens}, \citenamefont {Hone},\ and\ \citenamefont
  {Englund}}]{Shiue2015}%
  \BibitemOpen
  \bibfield  {author} {\bibinfo {author} {\bibfnamefont {R.J.}\ \bibnamefont
  {Shiue}}, \bibinfo {author} {\bibfnamefont {Y.}~\bibnamefont {Gao}}, \bibinfo
  {author} {\bibfnamefont {Y.}~\bibnamefont {Wang}}, \bibinfo {author}
  {\bibfnamefont {C.}~\bibnamefont {Peng}}, \bibinfo {author} {\bibfnamefont
  {A.D.}\ \bibnamefont {Robertson}}, \bibinfo {author} {\bibfnamefont {D.K.}\
  \bibnamefont {Efetov}}, \bibinfo {author} {\bibfnamefont {S.}~\bibnamefont
  {Assefa}}, \bibinfo {author} {\bibfnamefont {F.H.}\ \bibnamefont {Koppens}},
  \bibinfo {author} {\bibfnamefont {J.}~\bibnamefont {Hone}}, \ and\ \bibinfo
  {author} {\bibfnamefont {D.}~\bibnamefont {Englund}},\ }\bibfield  {title}
  {\enquote {\bibinfo {title} {{High-Responsivity Graphene–Boron Nitride
  Photodetector and Autocorrelator in a Silicon Photonic Integrated
  Circuit.}}}\ }\href@noop {} {\bibfield  {journal} {\bibinfo  {journal} {Nano
  Lett.}\ }\textbf {\bibinfo {volume} {15}},\ \bibinfo {pages} {7288}
  (\bibinfo {year} {2015})}\BibitemShut {NoStop}%
\bibitem [{\citenamefont {McKitterick}\ \emph {et~al.}(2016)\citenamefont
  {McKitterick}, \citenamefont {Prober},\ and\ \citenamefont
  {Rooks}}]{McKitterick2016}%
  \BibitemOpen
  \bibfield  {author} {\bibinfo {author} {\bibfnamefont {C.~B.}\ \bibnamefont
  {McKitterick}}, \bibinfo {author} {\bibfnamefont {D.~E.}\ \bibnamefont
  {Prober}}, \ and\ \bibinfo {author} {\bibfnamefont {M.~J.}\ \bibnamefont
  {Rooks}},\ }\bibfield  {title} {\enquote {\bibinfo {title} {{Electron-phonon
  cooling in large monolayer graphene devices}},}\ }\href {\doibase
  10.1103/PhysRevB.93.075410} {\bibfield  {journal} {\bibinfo  {journal} {Phys.
  Rev. B}\ }\textbf {\bibinfo {volume} {93}},\ \bibinfo {pages} {075410}
  (\bibinfo {year} {2016})}\BibitemShut {NoStop}%
\bibitem [{\citenamefont {Wang}\ \emph {et~al.}(2010)\citenamefont {Wang},
  \citenamefont {Strait}, \citenamefont {George}, \citenamefont {Shivaraman},
  \citenamefont {Shields}, \citenamefont {Chandrashekhar}, \citenamefont
  {Hwang}, \citenamefont {Rana}, \citenamefont {G.}, \citenamefont
  {Ruiz-Vargas},\ and\ \citenamefont {Park}}]{Wang2010}%
  \BibitemOpen
  \bibfield  {author} {\bibinfo {author} {\bibfnamefont {H.}~\bibnamefont
  {Wang}}, \bibinfo {author} {\bibfnamefont {J.~H.}\ \bibnamefont {Strait}},
  \bibinfo {author} {\bibfnamefont {P.~A.}\ \bibnamefont {George}}, \bibinfo
  {author} {\bibfnamefont {S.}~\bibnamefont {Shivaraman}}, \bibinfo {author}
  {\bibfnamefont {V.~B.}\ \bibnamefont {Shields}}, \bibinfo {author}
  {\bibfnamefont {M.}~\bibnamefont {Chandrashekhar}}, \bibinfo {author}
  {\bibfnamefont {J.}~\bibnamefont {Hwang}}, \bibinfo {author} {\bibfnamefont
  {F.}~\bibnamefont {Rana}}, \bibinfo {author} {\bibfnamefont {Spencer~M.}\
  \bibnamefont {G.}}, \bibinfo {author} {\bibfnamefont {C.~S.}\ \bibnamefont
  {Ruiz-Vargas}}, \ and\ \bibinfo {author} {\bibfnamefont {J.}~\bibnamefont
  {Park}},\ }\bibfield  {title} {\enquote {\bibinfo {title} {{Ultrafast
  relaxation dynamics of hot optical phonons in graphene}},}\ }\href@noop {}
  {\bibfield  {journal} {\bibinfo  {journal} {Appl. Phys. Lett.}\ }\textbf
  {\bibinfo {volume} {96}},\ \bibinfo {pages} {081917} (\bibinfo {year}
  {2010})}\BibitemShut {NoStop}%
\bibitem [{\citenamefont {Winnerl}\ \emph {et~al.}(2011)\citenamefont
  {Winnerl}, \citenamefont {Orlita}, \citenamefont {Plochocka}, \citenamefont
  {Kossacki}, \citenamefont {Potemski}, \citenamefont {Winzer}, \citenamefont
  {Malic}, \citenamefont {Knorr}, \citenamefont {Sprinkle}, \citenamefont
  {Berger}, \citenamefont {de~Heer}, \citenamefont {Schneider},\ and\
  \citenamefont {Helm}}]{Winnerl2011}%
  \BibitemOpen
  \bibfield  {author} {\bibinfo {author} {\bibfnamefont {S.}~\bibnamefont
  {Winnerl}}, \bibinfo {author} {\bibfnamefont {M.}~\bibnamefont {Orlita}},
  \bibinfo {author} {\bibfnamefont {P.}~\bibnamefont {Plochocka}}, \bibinfo
  {author} {\bibfnamefont {P.}~\bibnamefont {Kossacki}}, \bibinfo {author}
  {\bibfnamefont {M.}~\bibnamefont {Potemski}}, \bibinfo {author}
  {\bibfnamefont {T.}~\bibnamefont {Winzer}}, \bibinfo {author} {\bibfnamefont
  {E.}~\bibnamefont {Malic}}, \bibinfo {author} {\bibfnamefont
  {A.}~\bibnamefont {Knorr}}, \bibinfo {author} {\bibfnamefont
  {M.}~\bibnamefont {Sprinkle}}, \bibinfo {author} {\bibfnamefont
  {C.}~\bibnamefont {Berger}}, \bibinfo {author} {\bibfnamefont {W.~A.}\
  \bibnamefont {de~Heer}}, \bibinfo {author} {\bibfnamefont {H.}~\bibnamefont
  {Schneider}}, \ and\ \bibinfo {author} {\bibfnamefont {M.}~\bibnamefont
  {Helm}},\ }\bibfield  {title} {\enquote {\bibinfo {title} {{Carrier
  Relaxation in Epitaxial Graphene Photoexcited Near the Dirac Point}},}\
  }\href {\doibase 10.1103/PhysRevLett.107.237401} {\bibfield  {journal}
  {\bibinfo  {journal} {Phys. Rev. Lett.}\ }\textbf {\bibinfo {volume} {107}},\
  \bibinfo {pages} {237401} (\bibinfo {year} {2011})}\BibitemShut {NoStop}%
\bibitem [{\citenamefont {Tielrooij}\ \emph {et~al.}(2013)\citenamefont
  {Tielrooij}, \citenamefont {Song}, \citenamefont {Jensen}, \citenamefont
  {Centeno}, \citenamefont {Pesquera}, \citenamefont {Elorza}, \citenamefont
  {Bonn}, \citenamefont {Levitov},\ and\ \citenamefont
  {Koppens}}]{Tielrooij2013}%
  \BibitemOpen
  \bibfield  {author} {\bibinfo {author} {\bibfnamefont {K.~J.}\ \bibnamefont
  {Tielrooij}}, \bibinfo {author} {\bibfnamefont {J.~C.~W.}\ \bibnamefont
  {Song}}, \bibinfo {author} {\bibfnamefont {S.~A.}\ \bibnamefont {Jensen}},
  \bibinfo {author} {\bibfnamefont {A.}~\bibnamefont {Centeno}}, \bibinfo
  {author} {\bibfnamefont {A.}~\bibnamefont {Pesquera}}, \bibinfo {author}
  {\bibfnamefont {A.~Zurutuza}\ \bibnamefont {Elorza}}, \bibinfo {author}
  {\bibfnamefont {M.}~\bibnamefont {Bonn}}, \bibinfo {author} {\bibfnamefont
  {L.~S.}\ \bibnamefont {Levitov}}, \ and\ \bibinfo {author} {\bibfnamefont
  {F.~H.~L.}\ \bibnamefont {Koppens}},\ }\bibfield  {title} {\enquote {\bibinfo
  {title} {{Photoexcitation cascade and multiple hot-carrier generation in
  graphene}},}\ }\href@noop {} {\bibfield  {journal} {\bibinfo  {journal} {Nat.
  Phys.}\ }\textbf {\bibinfo {volume} {9}},\ \bibinfo {pages} {248}
  (\bibinfo {year} {2013})}\BibitemShut {NoStop}%
\bibitem [{\citenamefont {Lee}\ \emph {et~al.}(2008)\citenamefont {Lee},
  \citenamefont {Balasubramanian}, \citenamefont {Weitz}, \citenamefont
  {Burghard},\ and\ \citenamefont {Kern}}]{Lee2008}%
  \BibitemOpen
  \bibfield  {author} {\bibinfo {author} {\bibfnamefont {E.J.}\ \bibnamefont
  {Lee}}, \bibinfo {author} {\bibfnamefont {K.}~\bibnamefont
  {Balasubramanian}}, \bibinfo {author} {\bibfnamefont {R.T.}\ \bibnamefont
  {Weitz}}, \bibinfo {author} {\bibfnamefont {M.}~\bibnamefont {Burghard}}, \
  and\ \bibinfo {author} {\bibfnamefont {K.}~\bibnamefont {Kern}},\ }\bibfield
  {title} {\enquote {\bibinfo {title} {{ Contact and edge effects in graphene
  devices.}}}\ }\href@noop {} {\bibfield  {journal} {\bibinfo  {journal} {Nat.
  Nanotechnol.}\ }\textbf {\bibinfo {volume} {3}},\ \bibinfo {pages} {486}
  (\bibinfo {year} {2008})}\BibitemShut {NoStop}%
\bibitem [{\citenamefont {Xia}\ \emph {et~al.}(2009)\citenamefont {Xia},
  \citenamefont {Mueller}, \citenamefont {Golizadeh-Mojarad}, \citenamefont
  {Freitag}, \citenamefont {Lin}, \citenamefont {Tsang}, \citenamefont
  {Perebeinos},\ and\ \citenamefont {Avouris}}]{Xia2009}%
  \BibitemOpen
  \bibfield  {author} {\bibinfo {author} {\bibfnamefont {F.}~\bibnamefont
  {Xia}}, \bibinfo {author} {\bibfnamefont {T.}~\bibnamefont {Mueller}},
  \bibinfo {author} {\bibfnamefont {R.}~\bibnamefont {Golizadeh-Mojarad}},
  \bibinfo {author} {\bibfnamefont {M.}~\bibnamefont {Freitag}}, \bibinfo
  {author} {\bibfnamefont {Y.M.}\ \bibnamefont {Lin}}, \bibinfo {author}
  {\bibfnamefont {J.}~\bibnamefont {Tsang}}, \bibinfo {author} {\bibfnamefont
  {V.}~\bibnamefont {Perebeinos}}, \ and\ \bibinfo {author} {\bibfnamefont
  {P.}~\bibnamefont {Avouris}},\ }\bibfield  {title} {\enquote {\bibinfo
  {title} {{Photocurrent imaging and efficient photon detection in a graphene
  transistor.}}}\ }\href@noop {} {\bibfield  {journal} {\bibinfo  {journal}
  {Nano Lett.}\ }\textbf {\bibinfo {volume} {9}},\ \bibinfo {pages}
  {1039} (\bibinfo {year} {2009})}\BibitemShut {NoStop}%
\bibitem [{\citenamefont {Dean}\ \emph {et~al.}(2010)\citenamefont {Dean},
  \citenamefont {Young}, \citenamefont {Meric}, \citenamefont {Lee},
  \citenamefont {Wang}, \citenamefont {Sorgenfrei}, \citenamefont {Watanabe},
  \citenamefont {Taniguchi}, \citenamefont {Kim}, \citenamefont {Shepard},\
  and\ \citenamefont {Hone}}]{Dean2010}%
  \BibitemOpen
  \bibfield  {author} {\bibinfo {author} {\bibfnamefont {C.~R.}\ \bibnamefont
  {Dean}}, \bibinfo {author} {\bibfnamefont {A.~F.}\ \bibnamefont {Young}},
  \bibinfo {author} {\bibfnamefont {I.}~\bibnamefont {Meric}}, \bibinfo
  {author} {\bibfnamefont {C.}~\bibnamefont {Lee}}, \bibinfo {author}
  {\bibfnamefont {L.}~\bibnamefont {Wang}}, \bibinfo {author} {\bibfnamefont
  {S.}~\bibnamefont {Sorgenfrei}}, \bibinfo {author} {\bibfnamefont
  {K.}~\bibnamefont {Watanabe}}, \bibinfo {author} {\bibfnamefont
  {T.}~\bibnamefont {Taniguchi}}, \bibinfo {author} {\bibfnamefont
  {P.}~\bibnamefont {Kim}}, \bibinfo {author} {\bibfnamefont {K.~L.}\
  \bibnamefont {Shepard}}, \ and\ \bibinfo {author} {\bibfnamefont
  {J.}~\bibnamefont {Hone}},\ }\bibfield  {title} {\enquote {\bibinfo {title}
  {{Boron nitride substrates for high-quality graphene electronics.}}}\
  }\href@noop {} {\bibfield  {journal} {\bibinfo  {journal} {Nat.
  Nanotechnol.}\ }\textbf {\bibinfo {volume} {5}},\ \bibinfo {pages} {722}
  (\bibinfo {year} {2010})}\BibitemShut {NoStop}%
\bibitem [{\citenamefont {Wang}\ \emph {et~al.}(2013)\citenamefont {Wang},
  \citenamefont {Meric}, \citenamefont {Huang}, \citenamefont {Gao},
  \citenamefont {Gao}, \citenamefont {Tran}, \citenamefont {Taniguchi},
  \citenamefont {Watanabe}, \citenamefont {Campos}, \citenamefont {Muller},
  \citenamefont {Guo}, \citenamefont {Kim}, \citenamefont {Hone}, \citenamefont
  {Shepard},\ and\ \citenamefont {Dean}}]{Wang2013}%
  \BibitemOpen
  \bibfield  {author} {\bibinfo {author} {\bibfnamefont {L.}~\bibnamefont
  {Wang}}, \bibinfo {author} {\bibfnamefont {I.}~\bibnamefont {Meric}},
  \bibinfo {author} {\bibfnamefont {P.~Y.}\ \bibnamefont {Huang}}, \bibinfo
  {author} {\bibfnamefont {Q.}~\bibnamefont {Gao}}, \bibinfo {author}
  {\bibfnamefont {Y.}~\bibnamefont {Gao}}, \bibinfo {author} {\bibfnamefont
  {H.}~\bibnamefont {Tran}}, \bibinfo {author} {\bibfnamefont {T.}~\bibnamefont
  {Taniguchi}}, \bibinfo {author} {\bibfnamefont {K.}~\bibnamefont {Watanabe}},
  \bibinfo {author} {\bibfnamefont {L.~M.}\ \bibnamefont {Campos}}, \bibinfo
  {author} {\bibfnamefont {D.~A.}\ \bibnamefont {Muller}}, \bibinfo {author}
  {\bibfnamefont {J.}~\bibnamefont {Guo}}, \bibinfo {author} {\bibfnamefont
  {P.}~\bibnamefont {Kim}}, \bibinfo {author} {\bibfnamefont {J.}~\bibnamefont
  {Hone}}, \bibinfo {author} {\bibfnamefont {K.~L.}\ \bibnamefont {Shepard}}, \
  and\ \bibinfo {author} {\bibfnamefont {C.~R.}\ \bibnamefont {Dean}},\
  }\bibfield  {title} {\enquote {\bibinfo {title} {{One-Dimensional Electrical
  Contact to a Two-Dimensional Material}},}\ }\href@noop {} {\bibfield
  {journal} {\bibinfo  {journal} {Science}\ }\textbf {\bibinfo {volume}
  {342}},\ \bibinfo {pages} {614} (\bibinfo {year} {2013})}\BibitemShut
  {NoStop}%
\bibitem [{\citenamefont {Viljas}\ and\ \citenamefont
  {Heikkil\"a}(2010)}]{Viljas2010}%
  \BibitemOpen
  \bibfield  {author} {\bibinfo {author} {\bibfnamefont {J.~K.}\ \bibnamefont
  {Viljas}}\ and\ \bibinfo {author} {\bibfnamefont {T.~T.}\ \bibnamefont
  {Heikkil\"a}},\ }\bibfield  {title} {\enquote {\bibinfo {title}
  {{Electron-phonon heat transfer in monolayer and bilayer graphene}},}\ }\href
  {\doibase 10.1103/PhysRevB.81.245404} {\bibfield  {journal} {\bibinfo
  {journal} {Phys. Rev. B}\ }\textbf {\bibinfo {volume} {81}},\ \bibinfo
  {pages} {245404} (\bibinfo {year} {2010})}\BibitemShut {NoStop}%
\bibitem [{\citenamefont {Das~Sarma}\ \emph {et~al.}(2011)\citenamefont
  {Das~Sarma}, \citenamefont {Adam}, \citenamefont {Hwang},\ and\ \citenamefont
  {Rossi}}]{DasSarma2011}%
  \BibitemOpen
  \bibfield  {author} {\bibinfo {author} {\bibfnamefont {S.}~\bibnamefont
  {Das~Sarma}}, \bibinfo {author} {\bibfnamefont {S.}~\bibnamefont {Adam}},
  \bibinfo {author} {\bibfnamefont {E.~H.}\ \bibnamefont {Hwang}}, \ and\
  \bibinfo {author} {\bibfnamefont {E.}~\bibnamefont {Rossi}},\ }\bibfield
  {title} {\enquote {\bibinfo {title} {{Electronic transport in two-dimensional
  graphene}},}\ }\href {\doibase 10.1103/RevModPhys.83.407} {\bibfield
  {journal} {\bibinfo  {journal} {Rev. Mod. Phys.}\ }\textbf {\bibinfo {volume}
  {83}},\ \bibinfo {pages} {407} (\bibinfo {year} {2011})}\BibitemShut
  {NoStop}%
\bibitem [{\citenamefont {Low}\ \emph {et~al.}(2012)\citenamefont {Low},
  \citenamefont {Perebeinos}, \citenamefont {Kim}, \citenamefont {Freitag},\
  and\ \citenamefont {Avouris}}]{Low2012}%
  \BibitemOpen
  \bibfield  {author} {\bibinfo {author} {\bibfnamefont {T.}~\bibnamefont
  {Low}}, \bibinfo {author} {\bibfnamefont {V.}~\bibnamefont {Perebeinos}},
  \bibinfo {author} {\bibfnamefont {R.}~\bibnamefont {Kim}}, \bibinfo {author}
  {\bibfnamefont {M.}~\bibnamefont {Freitag}}, \ and\ \bibinfo {author}
  {\bibfnamefont {P.}~\bibnamefont {Avouris}},\ }\bibfield  {title} {\enquote
  {\bibinfo {title} {{Cooling of photoexcited carriers in graphene by internal
  and substrate phonons}},}\ }\href {\doibase 10.1103/PhysRevB.86.045413}
  {\bibfield  {journal} {\bibinfo  {journal} {Phys. Rev. B}\ }\textbf {\bibinfo
  {volume} {86}},\ \bibinfo {pages} {045413} (\bibinfo {year}
  {2012})}\BibitemShut {NoStop}%
\bibitem [{\citenamefont {Hwang}\ and\ \citenamefont
  {Das~Sarma}(2013)}]{Hwang2013}%
  \BibitemOpen
  \bibfield  {author} {\bibinfo {author} {\bibfnamefont {E.~H.}\ \bibnamefont
  {Hwang}}\ and\ \bibinfo {author} {\bibfnamefont {S.}~\bibnamefont
  {Das~Sarma}},\ }\bibfield  {title} {\enquote {\bibinfo {title} {{Surface
  polar optical phonon interaction induced many-body effects and hot-electron
  relaxation in graphene}},}\ }\href {\doibase 10.1103/PhysRevB.87.115432}
  {\bibfield  {journal} {\bibinfo  {journal} {Phys. Rev. B}\ }\textbf {\bibinfo
  {volume} {87}},\ \bibinfo {pages} {115432} (\bibinfo {year}
  {2013})}\BibitemShut {NoStop}%
\bibitem [{\citenamefont {Principi}\ \emph {et~al.}(2016)\citenamefont
  {Principi}, \citenamefont {Lundeberg}, \citenamefont {Hesp}, \citenamefont
  {Tielrooij}, \citenamefont {Koppens},\ and\ \citenamefont
  {Polini}}]{Principi2016}%
  \BibitemOpen
  \bibfield  {author} {\bibinfo {author} {\bibfnamefont {A.}~\bibnamefont
  {Principi}}, \bibinfo {author} {\bibfnamefont {M.~B.}\ \bibnamefont
  {Lundeberg}}, \bibinfo {author} {\bibfnamefont {N.C.H.}\ \bibnamefont
  {Hesp}}, \bibinfo {author} {\bibfnamefont {K.-J.}\ \bibnamefont {Tielrooij}},
  \bibinfo {author} {\bibfnamefont {F.H.L.}\ \bibnamefont {Koppens}}, \ and\
  \bibinfo {author} {\bibfnamefont {M.}~\bibnamefont {Polini}},\ }\href@noop {}
  {\enquote {\bibinfo {title} {{Super-Planckian electron cooling in a van der
  Waals stack}},}\ } \Eprint
  {http://arxiv.org/abs/1608.01516 [cond-mat.mes-hall]} {arXiv:1608.01516} \BibitemShut {NoStop}%
\bibitem [{\citenamefont {Freitag}\ \emph {et~al.}(2013)\citenamefont
  {Freitag}, \citenamefont {Low},\ and\ \citenamefont {Avouris}}]{Freitag2013}%
  \BibitemOpen
  \bibfield  {author} {\bibinfo {author} {\bibfnamefont {M.}~\bibnamefont
  {Freitag}}, \bibinfo {author} {\bibfnamefont {T.}~\bibnamefont {Low}}, \ and\
  \bibinfo {author} {\bibfnamefont {P.}~\bibnamefont {Avouris}},\ }\bibfield
  {title} {\enquote {\bibinfo {title} {{Increased Responsivity of Suspended
  Graphene Photodetectors}},}\ }\href {\doibase 10.1021/nl4001037} {\bibfield
  {journal} {\bibinfo  {journal} {Nano Lett.}\ }\textbf {\bibinfo {volume}
  {13}},\ \bibinfo {pages} {1644--1648} (\bibinfo {year} {2013})}\BibitemShut
  {NoStop}%
\bibitem [{\citenamefont {Song}\ \emph {et~al.}(2011)\citenamefont {Song},
  \citenamefont {Rudner}, \citenamefont {Marcus},\ and\ \citenamefont
  {Levitov}}]{Song2011}%
  \BibitemOpen
  \bibfield  {author} {\bibinfo {author} {\bibfnamefont {J.C.}\ \bibnamefont
  {Song}}, \bibinfo {author} {\bibfnamefont {M.S.}\ \bibnamefont {Rudner}},
  \bibinfo {author} {\bibfnamefont {C.M.}\ \bibnamefont {Marcus}}, \ and\
  \bibinfo {author} {\bibfnamefont {L.S.}\ \bibnamefont {Levitov}},\ }\bibfield
   {title} {\enquote {\bibinfo {title} {{Hot carrier transport and photocurrent
  response in graphene.}}}\ }\href@noop {} {\bibfield  {journal} {\bibinfo
  {journal} {Nano Lett.}\ }\textbf {\bibinfo {volume} {11}},\ \bibinfo {pages}
  {4688--4692} (\bibinfo {year} {2011})}\BibitemShut {NoStop}%
\bibitem [{\citenamefont {Duan}\ \emph {et~al.}(2016)\citenamefont {Duan},
  \citenamefont {Wang}, \citenamefont {Lai}, \citenamefont {Li}, \citenamefont
  {Watanabe}, \citenamefont {Taniguchi}, \citenamefont {Zebarjadi},\ and\
  \citenamefont {Andrei}}]{Duan2016}%
  \BibitemOpen
  \bibfield  {author} {\bibinfo {author} {\bibfnamefont {J.}~\bibnamefont
  {Duan}}, \bibinfo {author} {\bibfnamefont {X.}~\bibnamefont {Wang}}, \bibinfo
  {author} {\bibfnamefont {X.}~\bibnamefont {Lai}}, \bibinfo {author}
  {\bibfnamefont {G.}~\bibnamefont {Li}}, \bibinfo {author} {\bibfnamefont
  {K.}~\bibnamefont {Watanabe}}, \bibinfo {author} {\bibfnamefont
  {T.}~\bibnamefont {Taniguchi}}, \bibinfo {author} {\bibfnamefont
  {M.}~\bibnamefont {Zebarjadi}}, \ and\ \bibinfo {author} {\bibfnamefont
  {E.~Y.}\ \bibnamefont {Andrei}},\ }\href@noop {} {\enquote {\bibinfo {title}
  {{High Thermoelectric Power Factor in Graphene/hBN Devices}},}\ } (\bibinfo
  {year} {2016}),\ \Eprint {http://arxiv.org/abs/1607.00583
  [cond-mat.mes-hall]} {arXiv:1607.00583 [cond-mat.mes-hall]} \BibitemShut
  {NoStop}%
\bibitem [{\citenamefont {Hwang}\ \emph {et~al.}(2009)\citenamefont {Hwang},
  \citenamefont {Rossi},\ and\ \citenamefont {Das~Sarma}}]{Hwang2009}%
  \BibitemOpen
  \bibfield  {author} {\bibinfo {author} {\bibfnamefont {E.~H.}\ \bibnamefont
  {Hwang}}, \bibinfo {author} {\bibfnamefont {E.}~\bibnamefont {Rossi}}, \ and\
  \bibinfo {author} {\bibfnamefont {S.}~\bibnamefont {Das~Sarma}},\ }\bibfield
  {title} {\enquote {\bibinfo {title} {Theory of thermopower in two-dimensional
  graphene},}\ }\href {\doibase 10.1103/PhysRevB.80.235415} {\bibfield
  {journal} {\bibinfo  {journal} {Phys. Rev. B}\ }\textbf {\bibinfo {volume}
  {80}},\ \bibinfo {pages} {235415} (\bibinfo {year} {2009})}\BibitemShut
  {NoStop}%
\bibitem [{\citenamefont {Echtermeyer}\ \emph {et~al.}(2014)\citenamefont
  {Echtermeyer}, \citenamefont {Nene}, \citenamefont {Trushin}, \citenamefont
  {Gorbachev}, \citenamefont {Eiden}, \citenamefont {Milana}, \citenamefont
  {Sun}, \citenamefont {Schliemann}, \citenamefont {Lidorikis}, \citenamefont
  {Novoselov},\ and\ \citenamefont {Ferrari}}]{Echtermeyer2014}%
  \BibitemOpen
  \bibfield  {author} {\bibinfo {author} {\bibfnamefont {T.~J.}\ \bibnamefont
  {Echtermeyer}}, \bibinfo {author} {\bibfnamefont {P.~S.}\ \bibnamefont
  {Nene}}, \bibinfo {author} {\bibfnamefont {M.}~\bibnamefont {Trushin}},
  \bibinfo {author} {\bibfnamefont {R.~V.}\ \bibnamefont {Gorbachev}}, \bibinfo
  {author} {\bibfnamefont {A.~L.}\ \bibnamefont {Eiden}}, \bibinfo {author}
  {\bibfnamefont {S.}~\bibnamefont {Milana}}, \bibinfo {author} {\bibfnamefont
  {Z.}~\bibnamefont {Sun}}, \bibinfo {author} {\bibfnamefont {J.}~\bibnamefont
  {Schliemann}}, \bibinfo {author} {\bibfnamefont {E.}~\bibnamefont
  {Lidorikis}}, \bibinfo {author} {\bibfnamefont {K.~S.}\ \bibnamefont
  {Novoselov}}, \ and\ \bibinfo {author} {\bibfnamefont {A.~C.}\ \bibnamefont
  {Ferrari}},\ }\bibfield  {title} {\enquote {\bibinfo {title}
  {{Photothermoelectric and Photoelectric Contributions to Light Detection in
  Metal–Graphene–Metal Photodetectors}},}\ }\href {\doibase
  10.1021/nl5004762} {\bibfield  {journal} {\bibinfo  {journal} {Nano Lett.}\
  }\textbf {\bibinfo {volume} {14}},\ \bibinfo {pages} {3733} (\bibinfo
  {year} {2014})}\BibitemShut {NoStop}%
\bibitem [{\citenamefont {Tielrooij}\ \emph {et~al.}(2015)\citenamefont
  {Tielrooij}, \citenamefont {Massicotte}, \citenamefont {Piatkowski},
  \citenamefont {Woessner}, \citenamefont {Ma}, \citenamefont
  {Jarillo-Herrero}, \citenamefont {van Hulst},\ and\ \citenamefont
  {Koppens}}]{Tielrooij2015}%
  \BibitemOpen
  \bibfield  {author} {\bibinfo {author} {\bibfnamefont {K.~J.}\ \bibnamefont
  {Tielrooij}}, \bibinfo {author} {\bibfnamefont {M.}~\bibnamefont
  {Massicotte}}, \bibinfo {author} {\bibfnamefont {L.}~\bibnamefont
  {Piatkowski}}, \bibinfo {author} {\bibfnamefont {A.}~\bibnamefont
  {Woessner}}, \bibinfo {author} {\bibfnamefont {Q.}~\bibnamefont {Ma}},
  \bibinfo {author} {\bibfnamefont {P.}~\bibnamefont {Jarillo-Herrero}},
  \bibinfo {author} {\bibfnamefont {N.F.}\ \bibnamefont {van Hulst}}, \ and\
  \bibinfo {author} {\bibfnamefont {F.H.L.}\ \bibnamefont {Koppens}},\
  }\bibfield  {title} {\enquote {\bibinfo {title} {{Hot-carrier photocurrent
  effects at graphene–metal interfaces}},}\ }\href@noop {} {\bibfield
  {journal} {\bibinfo  {journal} {J. Phys. Condens. Matter}\ }\textbf {\bibinfo
  {volume} {27}},\ \bibinfo {pages} {164207} (\bibinfo {year}
  {2015})}\BibitemShut {NoStop}%
\bibitem []{SupplementaryContinuedReferences}%
  \BibitemOpen
  See Supplemental Material, which cites Refs.~\cite{Khomyakov2010,Mueller2009,Lui2010}, for details of thermal model, device fabrication, and electrical characterization
  \BibitemShut
  {NoStop}%
\bibitem [{\citenamefont {Khomyakov}\ \emph {et~al.}(2010)\citenamefont
  {Khomyakov}, \citenamefont {Starikov}, \citenamefont {Brocks},\ and\
  \citenamefont {Kelly}}]{Khomyakov2010}%
  \BibitemOpen
  \bibfield  {author} {\bibinfo {author} {\bibfnamefont {P.A.}\ \bibnamefont
  {Khomyakov}}, \bibinfo {author} {\bibfnamefont {A.A.}\ \bibnamefont
  {Starikov}}, \bibinfo {author} {\bibfnamefont {G.}~\bibnamefont {Brocks}}, \
  and\ \bibinfo {author} {\bibfnamefont {P.J.}\ \bibnamefont {Kelly}},\
  }\bibfield  {title} {\enquote {\bibinfo {title} {{Nonlinear screening of
  charges induced in graphene by metal contacts}},}\ }\href@noop {} {\bibfield
  {journal} {\bibinfo  {journal} {Phys. Rev. B}\ }\textbf {\bibinfo {volume}
  {82}},\ \bibinfo {pages} {115437} (\bibinfo {year} {2010})}\BibitemShut
  {NoStop}%
\bibitem [{\citenamefont {Mueller}\ \emph {et~al.}(2009)\citenamefont
  {Mueller}, \citenamefont {Xia}, \citenamefont {Freitag}, \citenamefont
  {Tsang},\ and\ \citenamefont {Avouris}}]{Mueller2009}%
  \BibitemOpen
  \bibfield  {author} {\bibinfo {author} {\bibfnamefont {T.}~\bibnamefont
  {Mueller}}, \bibinfo {author} {\bibfnamefont {F.}~\bibnamefont {Xia}},
  \bibinfo {author} {\bibfnamefont {M.}~\bibnamefont {Freitag}}, \bibinfo
  {author} {\bibfnamefont {J.}~\bibnamefont {Tsang}}, \ and\ \bibinfo {author}
  {\bibfnamefont {P.}~\bibnamefont {Avouris}},\ }\bibfield  {title} {\enquote
  {\bibinfo {title} {{Role of contacts in graphene transistors: A scanning
  photocurrent study.}}}\ }\href@noop {} {\bibfield  {journal} {\bibinfo
  {journal} {Phys. Rev. B}\ }\textbf {\bibinfo {volume} {79}},\ \bibinfo
  {pages} {245430} (\bibinfo {year} {2009})}\BibitemShut {NoStop}%
\bibitem [{\citenamefont {Lui}\ \emph {et~al.}(2010)\citenamefont {Lui},
  \citenamefont {Mak}, \citenamefont {Shan},\ and\ \citenamefont
  {Heinz}}]{Lui2010}%
  \BibitemOpen
  \bibfield  {author} {\bibinfo {author} {\bibfnamefont {C.~H.}\ \bibnamefont
  {Lui}}, \bibinfo {author} {\bibfnamefont {K.~F.}\ \bibnamefont {Mak}},
  \bibinfo {author} {\bibfnamefont {J.}~\bibnamefont {Shan}}, \ and\ \bibinfo
  {author} {\bibfnamefont {T.~F.}\ \bibnamefont {Heinz}},\ }\bibfield  {title}
  {\enquote {\bibinfo {title} {{Ultrafast Photoluminescence from Graphene}},}\
  }\href@noop {} {\bibfield  {journal} {\bibinfo  {journal} {Phys. Rev. Lett.}\
  }\textbf {\bibinfo {volume} {105}},\ \bibinfo {pages} {127404} (\bibinfo
  {year} {2010})}\BibitemShut {NoStop}%
\bibitem [{\citenamefont {Sun}\ \emph {et~al.}(2012)\citenamefont {Sun},
  \citenamefont {Aivazian}, \citenamefont {Jones}, \citenamefont {Ross},
  \citenamefont {Yao}, \citenamefont {Cobden},\ and\ \citenamefont
  {Xu}}]{Sun2012}%
  \BibitemOpen
  \bibfield  {author} {\bibinfo {author} {\bibfnamefont {D.}~\bibnamefont
  {Sun}}, \bibinfo {author} {\bibfnamefont {G.}~\bibnamefont {Aivazian}},
  \bibinfo {author} {\bibfnamefont {A.M.}\ \bibnamefont {Jones}}, \bibinfo
  {author} {\bibfnamefont {J.S.}\ \bibnamefont {Ross}}, \bibinfo {author}
  {\bibfnamefont {W.}~\bibnamefont {Yao}}, \bibinfo {author} {\bibfnamefont
  {D.}~\bibnamefont {Cobden}}, \ and\ \bibinfo {author} {\bibfnamefont
  {X.}~\bibnamefont {Xu}},\ }\bibfield  {title} {\enquote {\bibinfo {title}
  {{Ultrafast hot-carrier-dominated photocurrent in graphene.}}}\ }\href@noop
  {} {\bibfield  {journal} {\bibinfo  {journal} {Nat. Nanotechnol.}\ }\textbf
  {\bibinfo {volume} {7}},\ \bibinfo {pages} {114} (\bibinfo {year}
  {2012})}\BibitemShut {NoStop}%
\bibitem [{\citenamefont {Graham}\ \emph
  {et~al.}(2013{\natexlab{b}})\citenamefont {Graham}, \citenamefont {Shi},
  \citenamefont {Wang}, \citenamefont {Ralph}, \citenamefont {Park},\ and\
  \citenamefont {McEuen}}]{MWGraham2013}%
  \BibitemOpen
  \bibfield  {author} {\bibinfo {author} {\bibfnamefont {M.W.}\ \bibnamefont
  {Graham}}, \bibinfo {author} {\bibfnamefont {S.F.}\ \bibnamefont {Shi}},
  \bibinfo {author} {\bibfnamefont {Z.}~\bibnamefont {Wang}}, \bibinfo {author}
  {\bibfnamefont {D.C.}\ \bibnamefont {Ralph}}, \bibinfo {author}
  {\bibfnamefont {J.}~\bibnamefont {Park}}, \ and\ \bibinfo {author}
  {\bibfnamefont {P.L.}\ \bibnamefont {McEuen}},\ }\bibfield  {title} {\enquote
  {\bibinfo {title} {{Transient absorption and photocurrent microscopy show
  that hot electron supercollisions describe the rate-limiting relaxation step
  in graphene.}}}\ }\href@noop {} {\bibfield  {journal} {\bibinfo  {journal}
  {Nano Lett.}\ }\textbf {\bibinfo {volume} {13}},\ \bibinfo {pages}
  {5497} (\bibinfo {year} {2013}{\natexlab{b}})}\BibitemShut {NoStop}%
\bibitem [{\citenamefont {Jadidi}\ \emph {et~al.}(2016)\citenamefont {Jadidi},
  \citenamefont {Konig-Otto}, \citenamefont {Sushkov}, \citenamefont {Winnerl},
  \citenamefont {Drew}, \citenamefont {Murphy},\ and\ \citenamefont
  {Mittendorff}}]{Jadidi2016}%
  \BibitemOpen
  \bibfield  {author} {\bibinfo {author} {\bibfnamefont {M.~M.}\ \bibnamefont
  {Jadidi}}, \bibinfo {author} {\bibfnamefont {J.~C.}\ \bibnamefont
  {Konig-Otto}}, \bibinfo {author} {\bibfnamefont {A.~B.}\ \bibnamefont
  {Sushkov}}, \bibinfo {author} {\bibfnamefont {S.}~\bibnamefont {Winnerl}},
  \bibinfo {author} {\bibfnamefont {H.~D.}\ \bibnamefont {Drew}}, \bibinfo
  {author} {\bibfnamefont {T.~E.}\ \bibnamefont {Murphy}}, \ and\ \bibinfo
  {author} {\bibfnamefont {M.}~\bibnamefont {Mittendorff}},\ }\bibfield
  {title} {\enquote {\bibinfo {title} {{Nonlinear Terahertz Absorption of
  Graphene Plasmons}},}\ }\href@noop {} {\bibfield  {journal} {\bibinfo
  {journal} {Nano Lett.}\ }\textbf {\bibinfo {volume} {16}} (\bibinfo {year}
  {2016})}\BibitemShut {NoStop}%
\end{thebibliography}


\clearpage
\pagebreak
\onecolumngrid

\part*{Supplemental Material}

\renewcommand{\theequation}{S\arabic{equation}}
\renewcommand{\thesection}{S\arabic{section}}
\renewcommand{\thefigure}{S\arabic{figure}}
\setcounter{equation}{0}
\setcounter{section}{0}
\setcounter{figure}{0}

\section{Nonlinear Thermal Model}

The electron temperature in the graphene may be modeled by the following nonlinear differential equation:
\begin{equation}\label{eq:S1}
  \alpha T \frac{dT}{dt} + \beta_1 (T - T_L) + \beta_3 (T^3-T_L^3) = I(t)
\end{equation}
where $T$ represents the graphene electron temperature, $T_L$ is the lattice temperature, and $I(t)$ is the absorbed optical power per unit area.  $\alpha T$ is the specific heat in the graphene and the terms proportional to $\beta_1$ and $\beta_3$ describe momentum-conserving cooling and disorder-assisted supercollision cooling, respectively.

We re-write these equations in terms of $x \equiv T-T_L$, the deviation from the lattice temperature:
\begin{equation}\label{eq:S2}
  \alpha (T_L + x) \frac{dx}{dt} + \beta_1 x + \beta_3 \left[(T_L + x)^3 - T_L^3\right] = I(t)
\end{equation}
We next assume that $x \ll T_L$, i.e., the photoinduced change in electron temperature is small in comparison to the equilibrium (lattice) temperature.  With this assumption, $x(t)$ may be expanded in a power series in the intensity $I$,
\begin{equation}\label{eq:S3}
  x(t) = x^{(1)}(t) + x^{(2)}(t) + x^{(3)}(t) + \ldots
\end{equation}
Where $x^{(n)} \propto I^n$, and we are retaining terms up to third order. Substituting this expansion into Eq.~\eqref{eq:S2} gives
\begin{multline}\label{eq:S4}
  \alpha (T_L + x^{(1)} + x^{(2)} + x^{(3)}) \frac{d}{dt}(x^{(1)} + x^{(2)} + x^{(3)}) + \beta_1 (x^{(1)} + x^{(2)} + x^{(3)}) + \\ \beta_3 \left[(T_L + x^{(1)}+x^{(2)} + x^{(3)})^3 - T_L^3\right] = I(t)
\end{multline}
Next, we expand Eq.~\eqref{eq:S4} and separately equate the orders to obtain the following inhomogeneous linear differential equations for $x^{(1)}$, $x^{(2)}$ and $x^{(3)}$,
\begin{align}\label{eq:S5}
  \alpha T_L \frac{dx^{(1)}}{dt} + (\beta_1 + 3\beta_3 T_L^2) x^{(1)} &= I(t)
  \\\label{eq:S6}
  \alpha T_L \frac{dx^{(2)}}{dt} + (\beta_1 + 3\beta_3 T_L^2) x^{(2)} &=
    -\alpha x^{(1)}\frac{dx^{(1)}}{dt} - 3\beta_3 T_L \bigl[x^{(1)}\bigr]^2
  \\\label{eq:S7}
  \alpha T_L \frac{dx^{(3)}}{dt} + (\beta_1 + 3\beta_3 T_L^2)x^{(3)} &= - \alpha x^{(1)}\frac{dx^{(2)}}{dt} - \alpha x^{(2)}\frac{dx^{(1)}}{dt} - 6\beta_3 T_L x^{(1)}x^{(2)} - \beta_3 \bigl[x^{(1)}\bigr]^3
\end{align}
which can be re-written as:
\begin{align}\label{eq:S8}
  \frac{dx^{(1)}}{dt} + \gamma x^{(1)} &= \frac{I(t)}{\alpha T_L}
  \\\label{eq:S9}
  \frac{dx^{(2)}}{dt} + \gamma x^{(2)} &=
    -\frac{1}{T_L} x^{(1)}\frac{dx^{(1)}}{dt} - \frac{3\beta_3}{\alpha} \bigl[x^{(1)}\bigr]^2
  \\\label{eq:S10}
  \frac{dx^{(3)}}{dt} + \gamma x^{(3)} &=
    -\frac{1}{T_L} x^{(1)}\frac{dx^{(2)}}{dt} -\frac{1}{T_L} x^{(2)}\frac{dx^{(1)}}{dt} - \frac{6\beta_3}{\alpha} x^{(1)}x^{(2)} - \frac{\beta_3}{\alpha T_L} \bigl[x^{(1)}\bigr]^3
\end{align}
where
\begin{equation}\label{eq:S11}
  \gamma \equiv \frac{\beta_1 + 3\beta_3 T_L^2}{\alpha T_L}
\end{equation}
represents the equivalent (linearized) cooling rate, taking into account both cooling mechanisms.

For the two-laser illumination considered here, the optical intensity absorbed in the graphene is given by
\begin{equation}\label{eq:S12}
  I(t) = I_1 + I_2 + 2\sqrt{I_1I_2}\cos\Omega t
\end{equation}
where $I_1$ is the absorbed intensity of laser 1, $I_2$ is the absorbed intensity of laser 2, and $\Omega \equiv \omega_1 - \omega_2$ is the heterodyne beat frequency between the two lasers.

Substituting this expression into Eq.~\eqref{eq:S8}, one can find a solution for $x^{(1)}(t)$, which is used in turn to find $x^{(2)}(t)$ from Eq.~\eqref{eq:S9}, and $x^{(3)}(t)$ from Eq.~\eqref{eq:S10}.

The photovoltage produced through the Seebeck effect can be expressed as
\begin{equation}\label{eq:S13}
  V(t) = rT(T - T_L) = rx(x+T_L)
\end{equation}
where $rT$ is the Seebeck coefficient of graphene.  Substituting $x = x^{(1)}+x^{(2)}+x^{(3)} +\ldots$ into Eq.~\eqref{eq:S13}, evaluating only the dc component of $V(t)$, and retaining only terms up to the third order in $I$, one finds, after simplification:
\begin{align}
  \label{eq:S14}
  V(I_1,I_2) = r\biggl\{&\frac{I_1+I_2}{\alpha\gamma}
  +\beta_1\frac{(I_1+I_2)^2}{(\alpha\gamma T_L)^3}
  -(3\beta_3^2T_L^4 + 7 T_L^2\beta_1\beta_3)\frac{(I_1+I_2)^3}{T_L^6\alpha^5\gamma^5}\quad\ldots \\
  \label{eq:S15}
  +\:&2I_1I_2\biggl[\frac{\beta_1}{(\alpha\gamma T_L)^3}
  - (9T_L^4\beta_3^2 + 15T_L^2\beta_1\beta_3 + 2\beta_1^2) \frac{(I_1+I_2)}{T_L^6\alpha^5\gamma^5}\biggr] \frac{\gamma^2}{\Omega^2 + \gamma^2}\quad\ldots \\
  \label{eq:S16}
  -\:&2I_1I_2\biggl[(6T_L^2\beta_1\beta_3 - 2\beta_1^2) \frac{(I_1+I_2)}{T_L^6\alpha^5\gamma^5}\biggr] \left(\frac{\gamma^2}{\Omega^2 + \gamma^2}\right)^2
  \biggr\}
\end{align}
For the room-temperature conditions reported here ($T_L = 300$ K), we may make the additional approximation that $\beta_1 \ll \beta_3T_L^2$.  In this regime, the linearized cooling rate ($\gamma$) is determined primarily by supercollision cooling, even though both cooling processes contribute to the measured nonlinearity in the response.  With this assumption, Eqs.~\eqref{eq:S14}-\eqref{eq:S16} simplify to:
\begin{align}
  \label{eq:S17}
  V(I_1,I_2) = r\biggl\{&\frac{I_1+I_2}{\alpha\gamma}
  +\beta_1\frac{(I_1+I_2)^2}{(\alpha\gamma T_L)^3}
  -3\beta_3^2\frac{(I_1+I_2)^3}{T_L^2\alpha^5\gamma^5}\quad\ldots \\
  \label{eq:S18}
  +\:&2I_1I_2\biggl[\frac{\beta_1}{(\alpha\gamma T_L)^3}
  - 9 \beta_3^2 \frac{(I_1+I_2)}{T_L^2\alpha^5\gamma^5}\biggr] \frac{\gamma^2}{\Omega^2 + \gamma^2}\biggr\}
\end{align}
The photoinduced voltage can be rewritten as
\begin{align}
  \label{eq:S19}
  V(I_1,I_2) = &a_1 (I_1+I_2) + a_2 (I_1^2+I_2^2) - a_3 (I_1^3 + I_2^3) \quad\ldots \\
  \label{eq:S20}
    +\:&2a_2I_1I_2\biggl(1+\frac{\gamma^2}{\Omega^2 + \gamma^2}\biggr)
    - 3a_3I_1I_2(I_1+I_2)\biggl(1+\frac{2\gamma^2}{\Omega^2 + \gamma^2}\biggr)
\end{align}
where the coefficients $a_1$, $a_2$ and $a_3$ are given by
\begin{equation}\label{eq:S21}
  a_1 \equiv \frac{r}{\alpha\gamma},\quad a_2 \equiv \frac{\beta_1}{(\alpha\gamma T_L)^3},\quad a_3 \equiv \frac{3\beta_3^2}{T_L^2\alpha^5\gamma^5}
\end{equation}

\begin{figure}[htbp]
  \centering
  \includegraphics[scale=0.8]{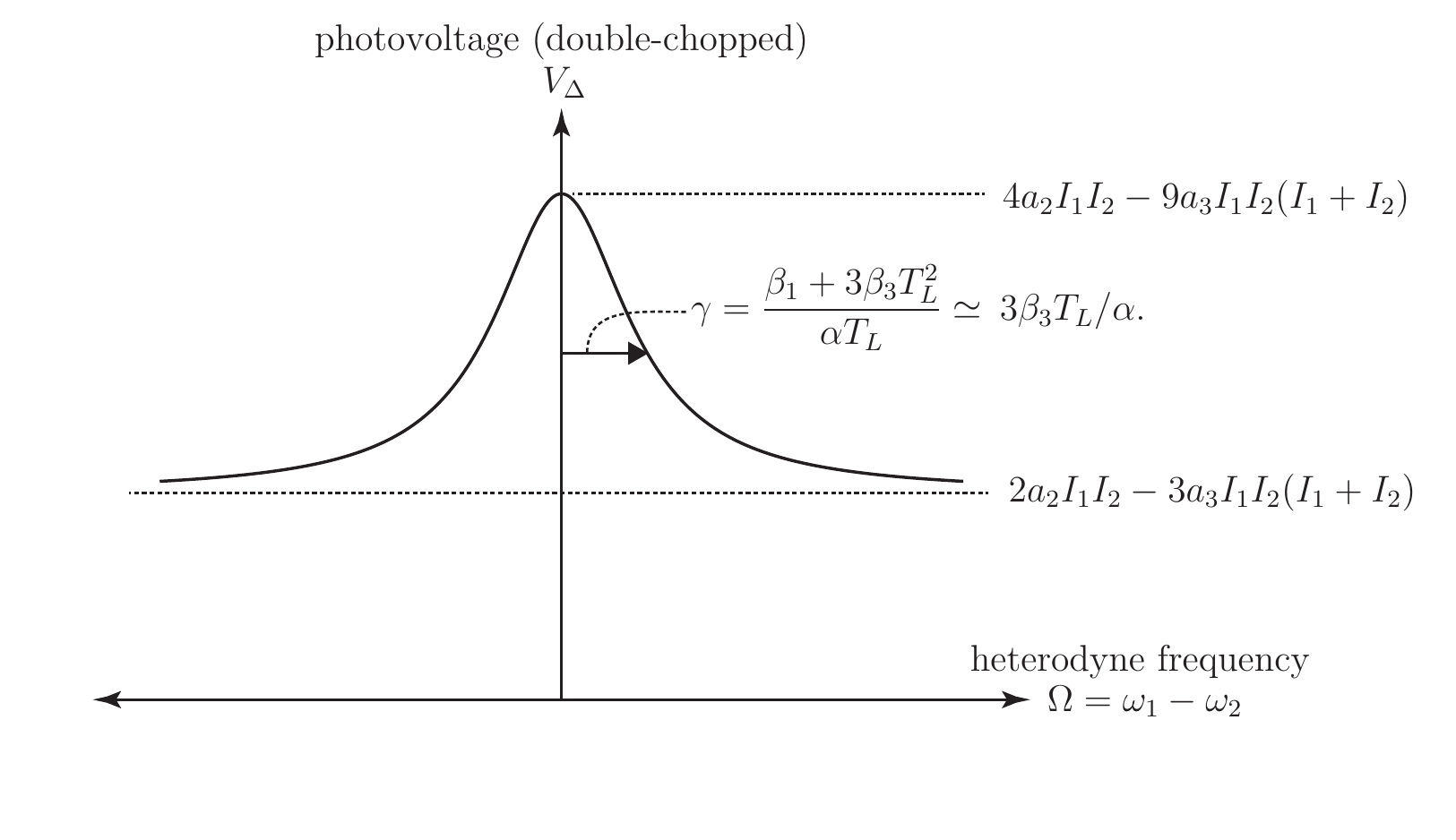}
  \caption{dc photovoltage $V_\Delta$ as a function of the heterodyne difference frequency $\Omega = \omega_1-\omega_2$.}\label{fig:S1}
\end{figure}

When the two beams $I_1$ and $I_2$ are double-chopped and synchronously detected at the chopper difference frequency, the lock-in amplifier produces a signal proportional to Eq.~\eqref{eq:S20}:
\begin{align}
  \label{eq:S22}
  V_{\Delta} &= V(I_1,I_2) - V(I_1,0) - V(0,I_2) + V(0,0) \\
  \label{eq:S23}
  &=\:2a_2I_1I_2\biggl[1+\frac{\gamma^2}{\Omega^2 + \gamma^2}\biggr]
    - 3a_3I_1I_2(I_1+I_2)\biggl[1+\frac{2\gamma^2}{\Omega^2 + \gamma^2}\biggr]
\end{align}
The dc photovoltage therefore has a Lorentzian dependence on the heterodyne difference frequency $\Omega \equiv \omega_1-\omega_2$, with a spectral width that is proportional to the carrier cooling rate $\gamma$, as shown schematically in Fig.~\ref{fig:S1}

\section{Nonlinear Photoresponse of the lower mobility sample }

\begin{figure}
  \centering
  \includegraphics[scale=1.25]{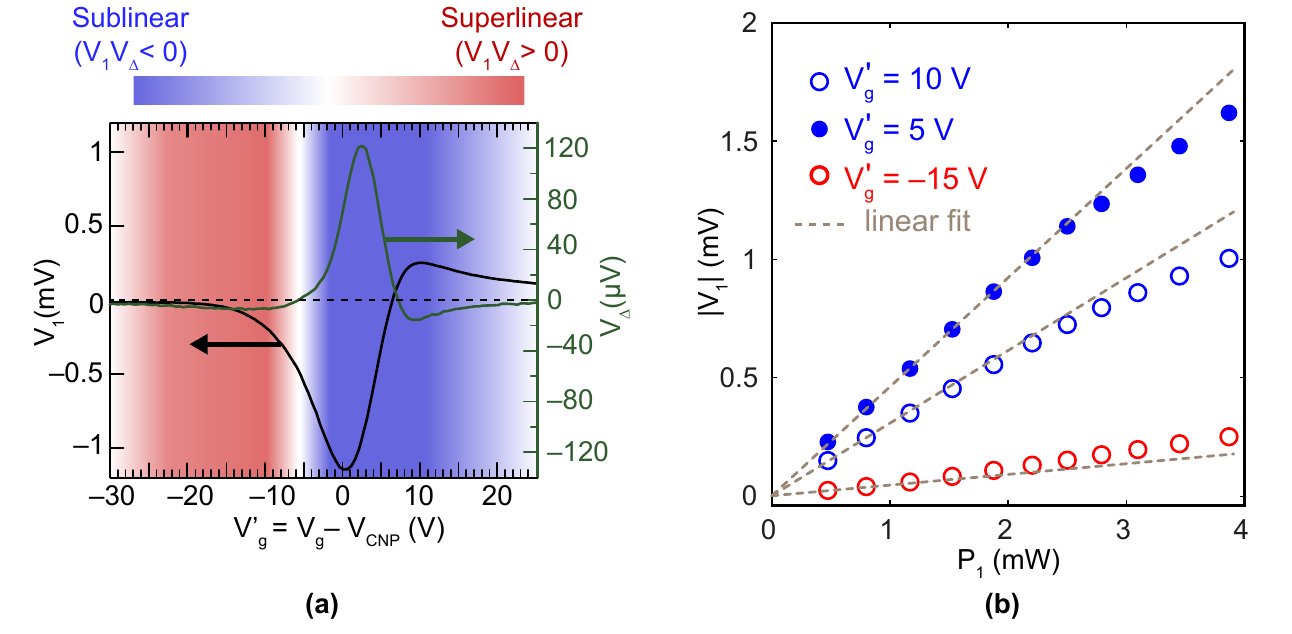}
  \caption{(a) Single-laser photovoltage ($V_1$), and nonlinear photomixing signal ($V_\Delta$) measured versus the gate voltage $V_g$, for the exfoliated graphene on SiO$_2$.  (b) Measured photovoltage $|V_1|$ versus optical input power, showing clearly the sublinear and superlinear behaviors.}
  \label{fig:S2}
\end{figure}

Fig.~\ref{fig:S2} shows a measurement similar to Fig.~2 performed on lower-mobility exfoliated graphene on SiO$_2$. In this device, the diffusion length is estimated to be only 500 nm, which is about one order of magnitude smaller than for the encapsulated device.  Because of this difference, the majority of the photoresponse in this device originates from the Fermi-level pinned region near the contact, where the carrier concentration is not as easily controlled by the applied gate voltage.  For positive gate voltages, Fermi level pinning produces a pn junction and charge-neutral region near the contact\cite{Khomyakov2010,Mueller2009}, which contributes to the observed sublinear response.  Otherwise, the response is qualitatively similar to that of the HBN-encapsulated device, and we observe a similar expected transition from supercollision cooling to conventional cooling under negative gate bias.

From the data in Fig.~\ref{fig:S2}, the sublinear-superlinear transition occurs at $V_g=-6$ V where the estimated Fermi level is $E_F = 80$ meV.  Assuming a disorder mean-free path of $l=40$ nm (which was independently determined from dc electrical measurements), we can use equation (5) in the main text to determine the ratio of the two rate coefficients, $\beta_1/\beta_3 = 5300$ $\rm{K}^2$.

\section{Linearized Cooling Rate at the Charge Neutral Point}

The thermal model presented in the manuscript and in Eq.~\eqref{eq:S1} ignores the fact that when the graphene is gated at the charge neutral point, the carriers are no longer degenerate, and under these conditions, the specific heat ($\alpha T$) and conventional cooling coefficient ($\beta_1$) must be modified to \cite{Viljas2010,Lui2010,McKitterick2016}:
\begin{align}
  \label{eq:S24}
  \alpha T &\rightarrow \alpha_2T^2,
  \quad\textrm{where}\quad \alpha_2 \equiv 18  \zeta(3) k_B^3 / \pi \hbar^2 V_F^2\\
  \label{eq:S25}
  \beta_1 &\rightarrow \beta_5 T^4,
  \quad\textrm{where}\quad \beta_5 \equiv 7 \pi^3 k_B^5 V_D^2/30 \rho \hbar^5 v_F^6
\end{align}
and the nonlinear thermal equation under these conditions becomes
\begin{equation}\label{eq:S26}
  \alpha_2 T^2 \frac{dT}{dt} + \beta_5 T^4 (T - T_L) + \beta_3 (T^3-T_L^3) = I(t)
\end{equation}

If Eq.~\eqref{eq:S26} is linearized about the lattice temperature, one obtains, analogous to Eq.~\eqref{eq:S5}
\begin{equation}\label{eq:S27}
  \alpha_2 T_L^2 \frac{dx}{dt} + (\beta_5 T_L^4 + \beta_3T_L^2)x = I(t)
\end{equation}
where $x = T-T_L$ is the photoinduced change in electron temperature relative to the lattice.  The linearized cooling rate is then
\begin{equation}\label{eq:S28}
  \gamma' = \frac{\beta_5 T_L^2 + 3\beta_3}{\alpha_2}
\end{equation}
which is shown by the red curve in Fig.~5b.

We expect that at low temperatures, $k_b T$ will be much smaller than $E_F^*$, the charge-puddle-limited Fermi level, in which case the cooling can instead be accurately described by Eq.~\eqref{eq:S1}.  The boundary between the two regimes can be estimated by equating Eq.~\eqref{eq:S11} and Eq.~\eqref{eq:S28}, which, for the parameters considered in Fig.~5 indicates that Eq.~\eqref{eq:S11} should only be applicable for $T_L<80$ K.  This condition is represented by the intersection between the blue and red curves in Fig.~5.

When the parameters determined from the low-temperature fit to Eq.~\eqref{eq:S11} are used in Eq.~\eqref{eq:S28}, with no additional free parameters, we correctly predict the observed cooling rate above 80 K, which further supports the model.

\section{Device Fabrication and DC Electrical Characterization}

Both devices considered here employed a doped silicon substrate ($\rho_{\rm Si} = 100$ $\Omega\cdot$cm), with 300 nm of thermally grown SiO$_2$ as a gate dielectric.  The substrate served as a large-area gate contact for adjusting the carrier concentration.

The HBN-encapsulated device \cite{Dean2010} was fabricated per the method described in\cite{Wang2013}. A piece of polypropylene carbonate (PPC) coated polydimethylsiloxane (PDMS) was first used to pick up HBN, monolayer graphene and another piece of HBN, in that order. The resulting heterostructure was then transferred to the aforementioned SiO$_2$ substrate, where electron beam lithography (EBL) was used to define a hydrogen silsesquioxane (HSQ) hard mask on poly(methy methacrylate) (PMMA). The surrounding areas were then etched in CHF$_3$ plasma to shape the device channel and expose the graphene edge. Afterwards, HSQ was lifted off and EBL was used again to define the contact leads and pads using PMMA, and 1.5 nm/20 nm/50 nm Cr/Pd/Au was e-beam evaporated and lifted off for edge contact. The HBN-encapsulated graphene channel length was 7 $\mu$m and width 0.7 $\mu$m .

For the second device, a single layer of graphene was mechanically exfoliated from bulk graphite and transferred directly to the SiO$_2$/Si substrate.  The exfoliated graphene exhibits a mobility about $\mu = 5,000$ cm$^2$V$^{-1}$s$^{-1}$, which was inferred from dc transport measurements.  Electron-beam lithography was used to pattern a bi-layer resist comprised of methyl methacrylate (MMA) and polymethy methacrylate (PMMA). The contacts were deposited using successive angled evaporations of chromium (15 nm) and gold (30 nm), thereby providing dissimilar contacts to the opposing edges of the graphene channel.  Dissimilar electrical contacts are not necessary when the optical beams are focused onto one contact, as for the measurements reported here, but this configuration also provides the thermal asymmetry needed for detection of spatially homogeneous or longer wavelength illumination.  The graphene channel length was 2.5 $\mu$m and width 7 $\mu$m.

\begin{figure}[htbp]
  \centering
  \includegraphics[scale=1.25]{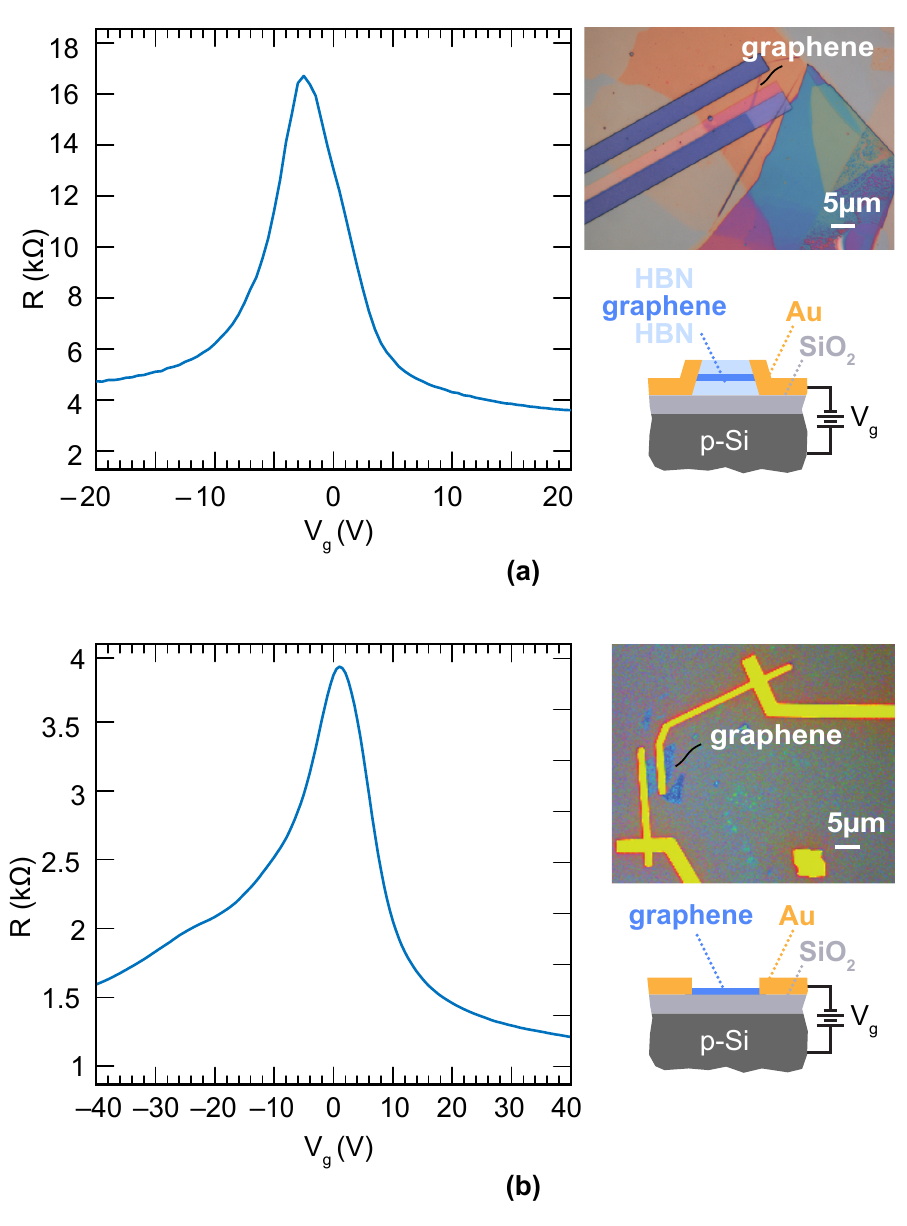}
  \caption{The dc resistance $R$ as a function of the applied gate voltage $V_g$ and the optical micrograph for (a) the HBN-encapsulated graphene device and (b) the exfoliated graphene on Si$\rm{O}_2$ device.  }\label{fig:S3}
\end{figure}

To quantify the electrical characteristics and gating behavior, we conducted unilluminated measurements of the dc resistance as a function of the gate voltage, for both the HBN-encapsulated device and the non-encapsulated device.  Fig.~\ref{fig:S3} shows the dc measurements, along with optical micrographs showing the graphene active region, contact geometry, and cross-sectional diagram.

\end{document}